\documentclass[twocolumn]{aastex62}

\usepackage{amsmath}
\usepackage{graphicx}
\usepackage{lipsum}
\usepackage{hyperref}
\usepackage{mathtools, nccmath}
\usepackage{longtable}
\usepackage{lineno}
\usepackage{enumitem}

\newcommand{\rev}{}
\newcommand{\RML}{}

\newcommand{\WT}{ZTF~SLRN-2020}

\graphicspath{{./}{figures/}}

\submitjournal{AAS Journals}

\shorttitle{JWST Observations of \WT}
\shortauthors{Lau et al.}

\begin{document}




\title{Revealing a main-sequence star that consumed a planet with JWST}



\correspondingauthor{Ryan M.\ Lau}
\email{ryan.lau@noirlab.edu}

\author[0000-0003-0778-0321]{Ryan M.\ Lau}
\affil{NSF NOIRLab, 950 N. Cherry Ave., Tucson, AZ 85719, USA}

\author[0000-0001-5754-4007]{Jacob E.\ Jencson}
\affil{IPAC, Mail Code 100-22, Caltech, 1200 E. California Blvd. Pasadena, CA 91125}

\author[0000-0003-3682-6632]{Colette Salyk}
\affil{Department of Physics and Astronomy, Vassar College, 124 Raymond Avenue, Poughkeepsie, NY 12604, USA }

\author[0000-0002-8989-0542]{Kishalay De}
\affil{MIT-Kavli Institute for Astrophysics and Space Research, 77 Massachusetts Ave., Cambridge, MA 02139, USA}

\author[0000-0003-2238-1572]{Ori D.\ Fox}
\affil{Space Telescope Science Institute, 3700 San Martin Drive, Baltimore, MD 21218, USA}

\author[0000-0001-9315-8437]{Matthew  J.\ Hankins}
\affil{Arkansas Tech University, 215 West O Street, Russellville, AR 72801, USA}

\author[0000-0002-5619-4938]{Mansi M.\ Kasliwal}
\affil{Division of Physics, Mathematics, and Astronomy, California Institute of Technology, Pasadena, CA 91125, USA}

\author[0000-0002-4834-369X]{Charles D.\ Keyes}
\affiliation{Space Telescope Science Institute, 3700 San Martin Drive, Baltimore, MD 21218, USA}

\author[0000-0002-1417-8024]{Morgan Macleod}
\affil{Center for Astrophysics | Harvard \& Smithsonian 60 Garden Street, MS-16, Cambridge, MA, 02138, USA}

\author[0000-0001-5644-8830]{Michael E.\ Ressler}
\affil{Jet Propulsion Laboratory, California Institute of Technology, MS 169-327, 4800 Oak Grove Drive, Pasadena, CA 91109}

\author[0000-0003-4725-4481]{Sam Rose}
\affil{Division of Physics, Mathematics, and Astronomy, California Institute of Technology, Pasadena, CA 91125, USA}





\begin{abstract}
The subluminous red nova (SLRN) \WT~is \rev{the most compelling} 
direct detection of a planet being consumed by its host star, a scenario known as a planetary engulfment event. We present JWST spectroscopy of \WT~taken +830 d after its optical emission peak using the NIRSpec fixed-slit $3-5$ $\mu$m high-resolution grating and the MIRI $5-12$ $\mu$m low-resolution spectrometer. 
NIRSpec reveals the $^{12}$CO fundamental band ($\nu=1-0$) in emission at $\sim4.7$ $\mu$m, Brackett-$\alpha$ emission\rev{, and the potential detection of PH$_3$ in emission at $\sim4.3$ $\mu$m.}
The JWST spectra are consistent with the claim that \WT~arose from a planetary engulfment event.
We utilize \texttt{DUSTY} to model the late-time $\sim1$--12 $\mu$m spectral energy distribution (SED) of \WT, where the best-fit parameters indicate the presence of warm, \rev{$720^{+80}_{-50}$ K}, circumstellar dust with a total dust mass of 
\rev{Log$\left(\frac{M_\mathrm{d}}{\mathrm{M}_\odot}\right)=-10.61^{+0.08}_{-0.16}$ M$_\odot$.}
We also fit a \texttt{DUSTY} model to archival photometry taken +320 d after peak that suggested the presence of a cooler, T$_\mathrm{d}=280^{+450}_{-20}$ K, and more massive, Log$\left(\frac{M_\mathrm{d}}{\mathrm{M}_\odot}\right)=-5.89^{+0.29}_{-3.21}$, circumstellar dust component. 
Assuming the cool component originates from the \WT~ejecta, we interpret the warm component as fallback from the ejecta.
From the late-time SED model we measure a luminosity of \rev{L$_* = 0.29^{+0.03}_{-0.06}$ L$_\odot$} for the remnant host star, which is consistent with a $\sim0.7$ M$_\odot$ K-type star that should not yet have evolved off the main sequence. 
If \WT~was not triggered by stellar evolution, we suggest that the planetary engulfment was due to orbital decay from tidal interactions between the planet and the host star.

\end{abstract}

\keywords{Circumstellar matter --- Stellar mergers --- Planetary system evolution --- \rev{Star-planet interactions}}

\section{Introduction}
\label{sec:intro}










The recent discovery of the subluminous red nova (SLRN) \WT~presented the \rev{most compelling} detection of a planetary engulfment event, where a $\lesssim10$ Jupiter-mass planet is believed to have been consumed by its Sun-like host star \citep{De2023,Soker2023}.
Such events provide unique insight into the evolution of planetary systems, especially for the population of short-orbital-period ($\lesssim10$ days) ``hot Jupiters'' \citep{Dawson2018}.
In particular, engulfment events capture the end stage of a planet influenced by dynamical processes (e.g.~\citealt{Rasio1996,Chatterjee2008}) and/or the post-main sequence evolution of its host star \citep{Xu2017,Macleod2018a, Stephan2020,Grunblatt2023, Yarza2023}. Anomalous chemical signatures from stars have notably been associated with planetary engulfments due to the incorporation of planetary material in the outer layers of the star \citep{Laughlin1997,Pinsonneault2001,Spina2021, SF2021}. 
The IR-luminous observational signatures of planetary engulfment events also probe the dusty and self-obscured physics of stellar coalescence, an important \RML{mechanism in the formation and evolution of stars.} \RML{\citep{DeMarco2017,Kashi2017, Kashi2019, Matsumoto2022,Macleod2022,Wolf2024}}.


\WT~was initially identified as a nova-like, transient optical outburst by the Zwicky Transient Facility (ZTF; \citealt{Bellm2019}) and coincided with a luminous IR counterpart captured by the ongoing mid-IR survey from the Near-Earth Object Wide-field Infrared Survey Explorer (NEOWISE) Reactivation mission \citep{Mainzer2014}. 
Nova-like optical transients are notably common in the Galactic disk and typically attributed to dwarf novae, classical novae, or young stellar outbursts. However, a key characteristic of those scenarios are strong atomic emission lines indicative of hot gas, while spectroscopic observations of \WT~during its outburst revealed an almost featureless, red continuum \citep{De2023}. Molecular features from \WT~such as TiO, VO, and CO were later revealed by ground-based near-IR spectroscopic follow-up from instruments like TripleSpec on the 200-inch \rev{Hale Telescope at the Palomar Observatory (P200)} and the Near-Infrared Echellette Spectrometer (NIRES) on the Keck-II Telescope \citep{Wilson2004,Herter2008} that instead suggested the presence of a cool outer photosphere consistent with an M-type giant with a temperature of around 3600 K \citep{De2023}. These cool molecular features as well as the short-lived optical outburst and persistent IR emission exhibited by \WT\ are hallmarks of an emerging class of transients known as ``red novae,'' which are associated with merger of two stars \citep{Munari2002, Tylenda2011,Ivanova2013, Karambelkar2023}. 

A comparison of \WT's low optical luminosity, $\sim10^{35}$ erg s$^{-1}$, and $\lesssim100$ d outburst duration, both of which are proportional to the companion mass, against the luminosity and duration of red novae from stellar-mass mergers demonstrated \WT~is consistent with the merger of a star and planet-mass companion (See Fig.~1 from \citealt{De2023}). 
Similar to the stellar-mass mergers, the optical outburst from \WT~was likely powered by hydrogen recombination in the ejected material while the IR-luminous aftermath arises from the formation of circumstellar dust \citep{Macleod2022,De2023}.
Observations of the \WT~progenitor from the United Kingdom Infrared Telescope (UKIRT) Galactic Plane survey \citep{Lawrence2007} provided constraints on the stellar component as a Sun-like, 0.8-1.5 M$_\odot$ star assuming a distance of 4 kpc \citep{De2023}.

The mechanism triggering the planetary engulfment event in \WT~is, however, uncertain since the progenitor mass range is consistent with a star that is either on or evolving off the main sequence. It is therefore unclear whether dynamical processes such as tidal decay of the orbit  and/or post-main sequence stellar evolution of the host star played the dominant role in the engulfment. The properties of circumstellar material from the ejecta of the planetary engulfment were also uncertain due to the lack of sensitive mid-IR follow-up capabilities at the time of the \WT~outburst. JWST's sensitive mid-IR spectroscopic capabilities are therefore ideal to probe the dusty aftermath of \WT~and investigate the open questions \RML{on planetary engulfment events}.

In this paper, we present late-time ($+830$ d) 3--12 $\mu$m spectroscopic follow-up observations with JWST and near-contemporaneous ground-based near-IR photometry from Gemini-North. The JWST and Gemini-N observations are described in Sec.~\ref{sec:Obs}. In Sec.~\ref{sec:results}, we identify notable features from the \WT~observations and \rev{investigate} the properties of the circumstellar material.
We also re-analyze earlier (+320 d) \WT~photometry from \citet{De2023} using \texttt{DUSTY} to investigate the evolving circumstellar dust properties. We then discuss the implications of the circumstellar and remnant host star properties and address the nature of \WT~in Sec.~\ref{sec:discussion}. 




\begin{figure}[t]
    \includegraphics[width=0.99\linewidth]{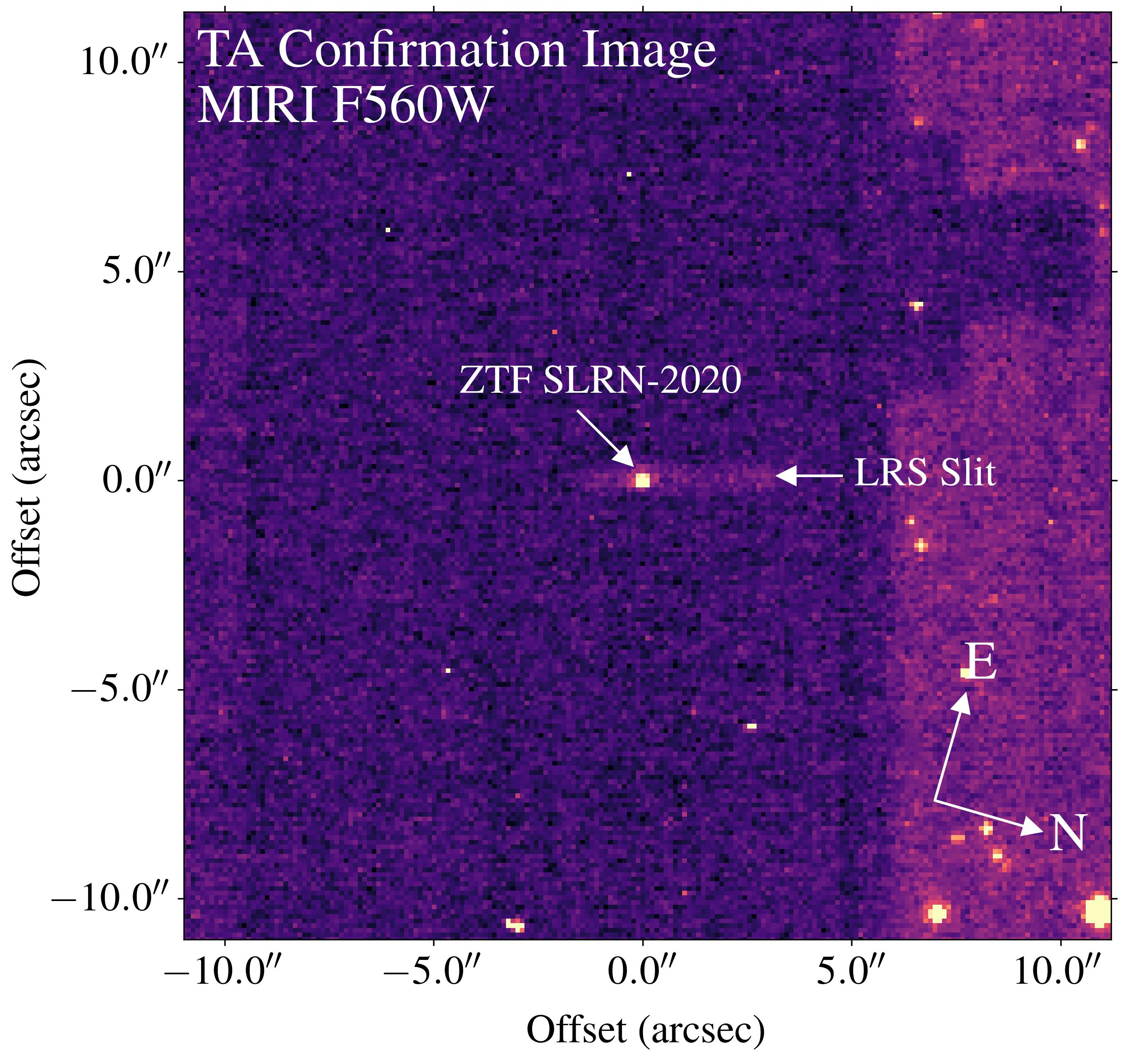}
    \caption{JWST/MIRI target acquisition (TA) confirmation image for the LRS observations of \WT~taken on 2022 September 5 with the F560W filter. \WT~is shown within the $4.7\times0.51$\arcsec~LRS slit.} 
    \label{fig:TA}
\end{figure}

\begin{deluxetable}{lp{1.3cm}p{1.8cm}p{2.4cm}}
\tablecaption{Observations of \WT}
\tablewidth{0pt}
\tablehead{Observatory & Obs.~Date (MJD) & Wavelength & Obs.~Mode}\
\startdata
Gemini-N/NIRI & 59809  & $J$ (1.25 $\mu$m), $H$ (1.65 $\mu$m) & Imaging \\
JWST/NIRSpec & 59827 & 3--5 $\mu$m & Fixed-Slit Spec. (R $\sim2700$) \\
JWST/MIRI & 59827 & 5--12 $\mu$m & Low-Resolution Spec. (R $\sim100$) \\
\enddata
\tablecomments{Summary of the $\sim1-12$ $\mu$m observations of \WT~presented in this work from Gemini-N/NIRI, JWST/NIRSpec, and JWST/MIRI.}
\label{tab:Obs}
\end{deluxetable}

\begin{figure*}[t]
    \includegraphics[width=0.98\linewidth]{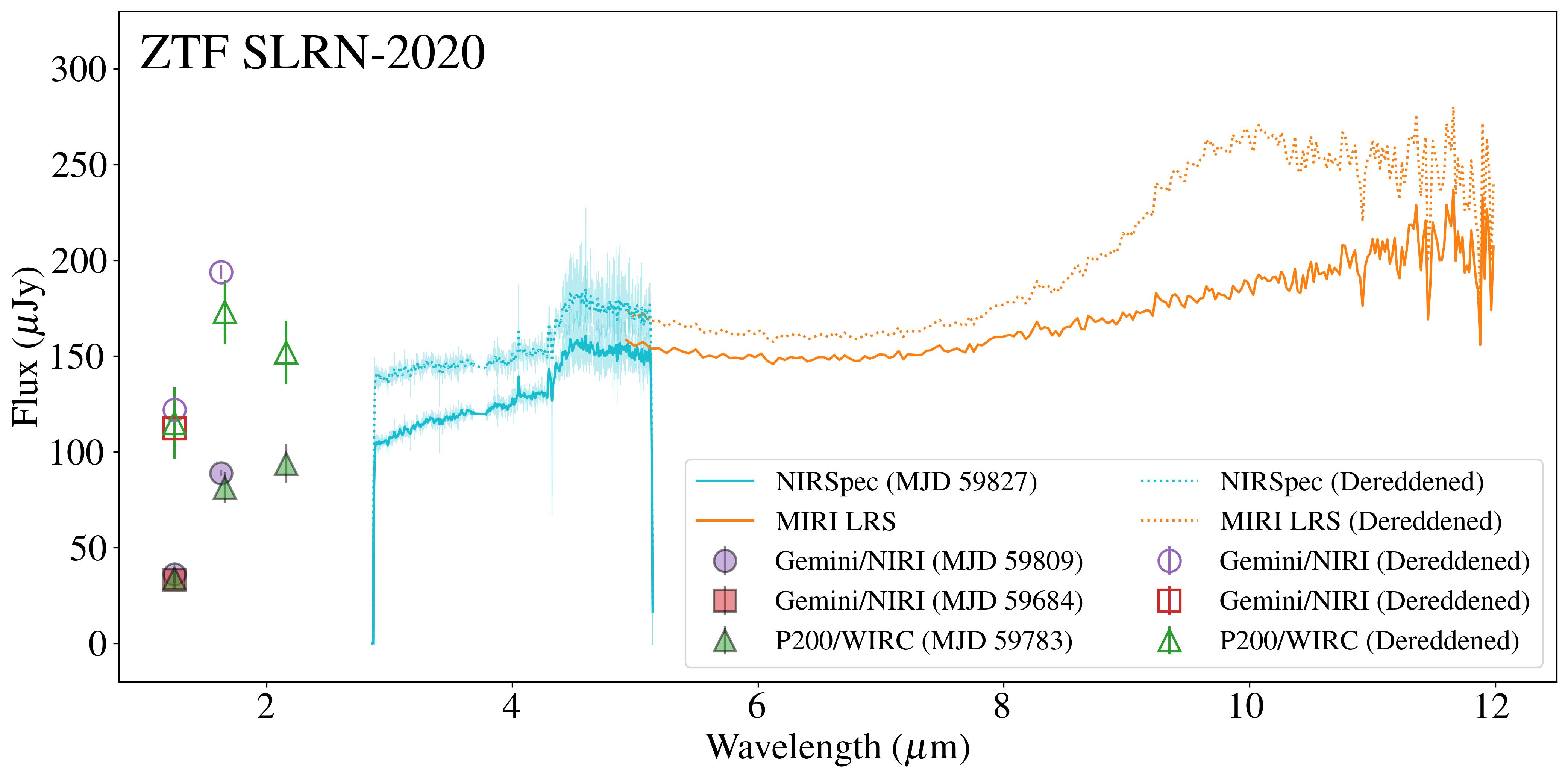}
    \includegraphics[width=0.98\linewidth]{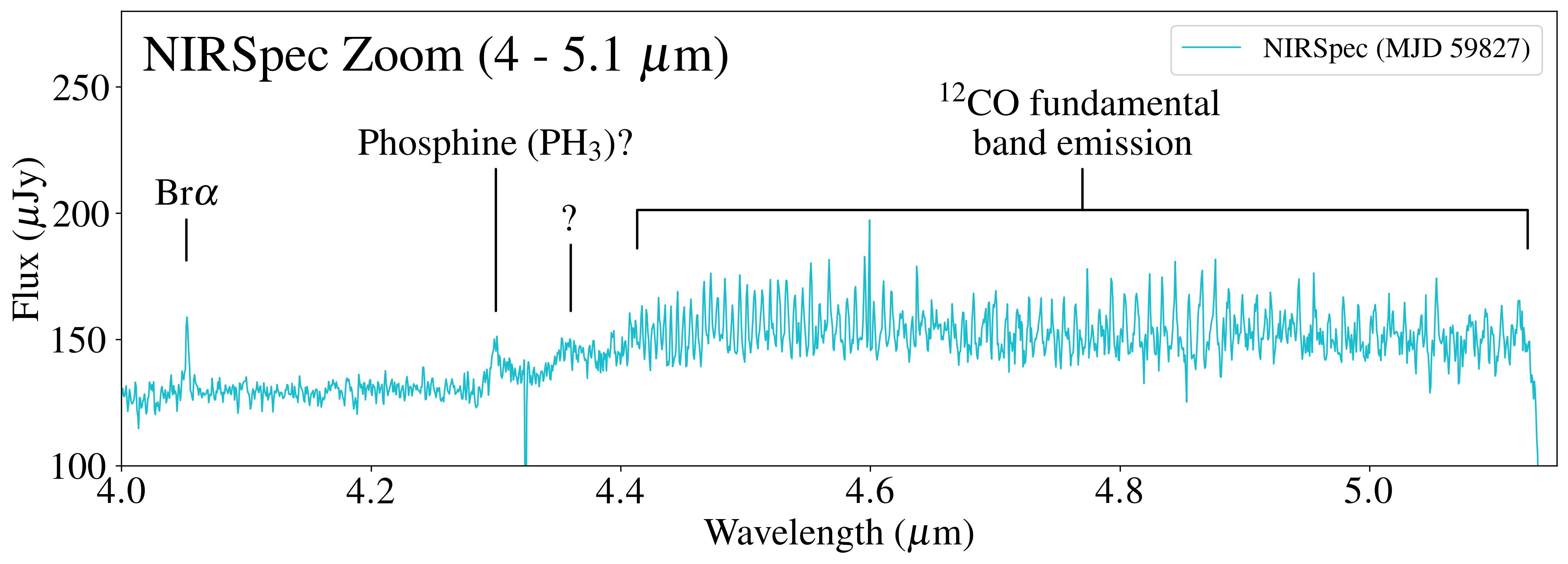}
    \caption{\rev{(\textit{Top})} Observed and dereddened+normalized spectral energy distribution of \WT~taken by Gemini/NIRI, P200/WIRC, JWST/NIRSpec, and JWST/MIRI LRS. NIRSpec and MIRI observations were taken on the same day (MJD 59827) in a non-interruptible sequence. The NIRSpec spectrum has been smoothed by convolving with a Gaussian with a 12-pixel FWHM and is overlaid on the non-convolved spectrum. Note that an additive normalization has been applied to the NIRSpec spectrum as described in Sec.~\ref{sec:norm}. \rev{(\textit{Bottom})} \rev{Observed NIRSpec spectrum overlaid with} notable spectroscopic \rev{features. The apparent narrow absorption ``feature'' to the right of the potential PH$_3$ feature is likely an artifact due to a bad pixel.}} 
    \label{fig:SED}
\end{figure*}

\section{Observations and Data Reduction}
\label{sec:Obs}

\subsection{JWST/NIRSpec Fixed Slit Spectroscopy}

\WT~was observed with JWST on 2022 September 5 in Cycle 1 as part of the program GTO 1240 (PI M.~Ressler). \WT~was one of three targets of opportunity observed in GTO 1240 and was selected based on the detection of its IR-luminous counterpart from NEOWISE \citep{De2023}.
Spectroscopic observations of \WT~covering $\sim3-12$ $\mu$m were obtained using NIRSpec Fixed Slit Spectroscopy and MIRI Low Resolution Spectroscopy in a non-interruptible sequence. 

NIRSpec observations utilized the $3.2\times0.2$\arcsec ~S200A1 slit with the G395H/F290LP Grating/Filter combination to cover a wavelength range of $\lambda=2.9-5.1$ $\mu$m with a resolving power of $\sim2700$. Note that the physical gap in the NIRSpec detectors results in a wavelength gap between 3.69 and 3.79 $\mu$m for the fixed-slit G395H/F290LP observations.
Target acquisition (TA) was performed on the same target as the science observation (i.e.~\WT) using the NIRSpec Wide Aperture TA (WATA) method with the SUB2048 subarray, F140X filter, and the NRSRAPID acquisition readout  pattern. The target acquisition image was inspected to ensure that \WT~was successfully acquired, which was important given its location in a crowded field with close nearby sources \citep{De2023}.

Observations of \WT~with NIRSpec were performed using the NRSIRS2RAPID readout pattern with the FULL subarray and two primary dither positions with no sub-pixel dithers. The total exposure time was 13.6 min with 28 groups per integration and two integrations (one per dither position). The NIRSpec observations were reduced using version \rev{1.16.0} of the JWST science calibration pipeline with 
the \rev{1298} CRDS context. The raw \texttt{uncal} files were downloaded from the Mikulski Archive for Space Telescopes (MAST)\footnote{\url{https://mast.stsci.edu/portal/Mashup/Clients/Mast/Portal.html}} and run through each stage of the pipeline.
\rev{The accuracy of the wavelength calibration for the high resolution ($\mathrm{R}\sim2700$) grating is $\sim15$ km s$^{-1}$, or $\sim1/8$ of a resolution element \citep{Boeker2023}.}

The \texttt{extract\_1d} step was re-run on the reduced level 3 NIRSpec \texttt{s2d} file \rev{using a 0.8\arcsec} aperture \rev{with \texttt{ystart} = 19 and \texttt{ystop} = 26. Since the observations were taken with two primary dither positions, no additional background subtraction was utilized in the \texttt{extract\_1d} step.} 
For the \texttt{DUSTY} spectral energy distribution analysis (Sec.~\ref{sec:DUSTY}), the spectrum was smoothed by convolving with a Gaussian kernel with a full width at half maximum (FWHM) of 12 pixels. Figure~\ref{fig:SED} presents the convolved and non-convolved NIRSpec spectra of \WT~as well as spectra that have been dereddened to correct for the estimated interstellar extinction ($A_V\approx3.6$ mag; \citealt{De2023}). A $2\%$ spectrophotometric uncertainty is adopted across the NIRSpec spectrum \citep{Boeker2023}. 
\RML{The} root-mean-square noise of the \RML{observed spectrum over the spectral pixels in the apparently featureless 3.0 - 4.0 $\mu$m wavelength range} is $5.5$ $\mu$Jy.

\subsection{JWST/MIRI Low Resolution Spectroscopy}
\label{sec:MIRI}

JWST/MIRI observations of \WT~with the low-resolution spectrometer (LRS) covered 5--12 $\mu$m with a resolving power of R $\sim100$ and were obtained on 2022 September 5 in a non-interruptible sequence with the NIRSpec observations. Due to the uncertain 5.6 $\mu$m emission from \WT~prior to the JWST observations, a nearby bright star was used for an offset target acquisition. The F560W TA confirmation image shown in Fig.~\ref{fig:TA} was taken to verify \WT~was successfully aligned in the $4.7\times0.51$\arcsec~LRS slit.
The MIRI LRS observations of \WT~were taken with the FULL subarray using the FASTR1 readout pattern and the 2-pt \texttt{ALONG SLIT NOD} dither type. The total MIRI LRS exposure time on \WT~was 13 min with 140 groups per integration/dither. 

The MIRI LRS data were reduced with version \rev{1.16.0} of the JWST science calibration pipeline 
with the \rev{1298} CRDS context. \rev{The default output from the \texttt{extract\_1d} step from the pipeline was used for the analysis in this work.}
The 4.9--12 $\mu$m range of the LRS spectrum is shown in Fig.~\ref{fig:SED} both before and after dereddening to correct for interstellar extinction. A $5\%$ spectrophotometric uncertainty is adopted across the LRS spectrum \citep{Wright2023}.

\subsection{NIRSpec-MIRI Normalization}
\label{sec:norm}

In the overlapping 4.9--5.1 $\mu$m wavelength range between the NIRSpec and MIRI spectra there is a \rev{6.0} $\mu$Jy ($4\%$) offset in the flux density. Figure~\ref{fig:SED} shows this offset, where the \rev{MIRI} spectrum is slightly \rev{brighter} than the \rev{NIRSpec} spectrum in the 4.9--5.1 $\mu$m wavelength range. In order to normalize the NIRSpec and MIRI spectra for spectral analysis and spectral energy distribution (SED) fitting, a \rev{negative} offset of \rev{6.0} $\mu$Jy was applied to the entire \rev{MIRI} spectrum. \rev{The normalization was applied to the MIRI spectrum since the 4\% difference from the overlapping NIRSpec range falls within the 5\%  spectrophotometric uncertainty \citep{Wright2023}} This normalized \rev{MIRI} spectrum is used for the data analysis. 
\subsection{Ground-based Near-IR Imaging}

We obtained high spatial resolution Adaptive Optics (AO) assisted imaging using the Near-Infrared Imager (NIRI) on the Gemini North telescope \citep{Hodapp2003}. The source was observed on UT 2022-08-18 as part of a Director's Discretionary Program (GN-2022A-DD-106; PI: K. De). We obtained dithered exposures of the field using Laser Guide Star (LGS) correction for a total exposure time of 60\,s, 60\,s and 120\,s in $J$, $H$ and $K$ bands respectively. The raw images were detrended and stacked using the \texttt{DRAGONS} pipeline \citep{Labrie2019}, using the source catalog from NIR images of the field presented in \citet{De2023} for astrometric and photometric calibration. We measured $J$ and $H$ (Vega) magnitudes of $J=19.08\pm0.04$ and $H=17.66\pm 0.02$, but we were unable to obtain a photometric solution for the $K$-band observations due to clouds.


\section{Results and Analysis}
\label{sec:results}



\subsection{Late-Time Infrared Emission from \WT}
\label{sec:latetimeir}

The late-time, ground-based and JWST observations of \WT~obtained $\gtrsim750$ d after its $i$-band peak (MJD 58993; \citealt{De2023}) are shown in Figure~\ref{fig:SED} and reveal a ``red" continuum rising from $\sim30$ to $\sim200$ $\mu$Jy between $\sim1-12$ $\mu$m.
Near-IR photometry from Gemini-N taken on 2022 August 18 (MJD 59809) is notably consistent with P200 ($J = 19.17 \pm 0.15$,
$H = 17.75 \pm 0.09$, and $Ks = 17.13 \pm 0.10$ mag) and previous Gemini-N ($J = 19.17 \pm 0.05$ mag) observations reported by \citet{De2023} that were taken on 2022 July 23 (MJD 59783) and 2022 April 15 (MJD 59684), respectively. The stability in the near-IR over the 4-month timescale suggests it probes emission from the photosphere of the remnant star as opposed to cooling circumstellar dust. The mid-IR $\gtrsim4$ $\mu$m emission that increases towards longer wavelengths likely traces thermal continuum emission from circumstellar dust. 

The dereddened IR emission from \WT~was corrected for interstellar extinction using the $R(V)=3.1$ model from the \texttt{dust\_extinction v1.3}\footnote{\url{https://dust-extinction.readthedocs.io/en/stable/}} python package \citep{Gordon2023} adopting a foreground extinction of A$_V=3.6$ mag \citep{De2023}.
Although the dereddened spectral energy distribution (SED) exhibits the 9.7 $\mu$m silicate emission feature, its presence is speculative because it appears only after dereddening the apparently featureless mid-IR continuum observed by the LRS. The shape and strength of the feature in the dereddened spectrum is therefore determined by the adopted A$_V$ and the profile of the silicate emissivity in the \citet{Gordon2023} extinction correction. 

The SED of \WT~exhibits the following notable features:

\begin{itemize}
\itemsep0em 
  \item Emission peak in the near-IR at $H$-band (1.64 $\mu$m)
  \item Detection of Br$\alpha$ ($\lambda = 4.05$ $\mu$m) emission
  \item \rev{Potential detection of phosphine (PH$_3$) emission at $\sim4.3$ $\mu$m}
  \item $^{12}$CO fundamental band ($\nu=1-0$) emission at $\sim4.7$ $\mu$m 
\end{itemize}


\noindent
Since thermal dust emission should not peak at such short wavelengths ($\lambda\sim1.5$ $\mu$m) due to the $\sim1000$ K sublimation temperature of dust grains, the presence of the near-IR peak in the dereddened spectrum supports the hypothesis that the near-IR emission traces the photosphere of the remnant star. The effective temperature and luminosity of the remnant star are investigated in more detail using \texttt{DUSTY} radiative transfer models in Sec.~\ref{sec:DUSTY}. 

\begin{figure}[t!]
    \includegraphics[width=0.98\linewidth]{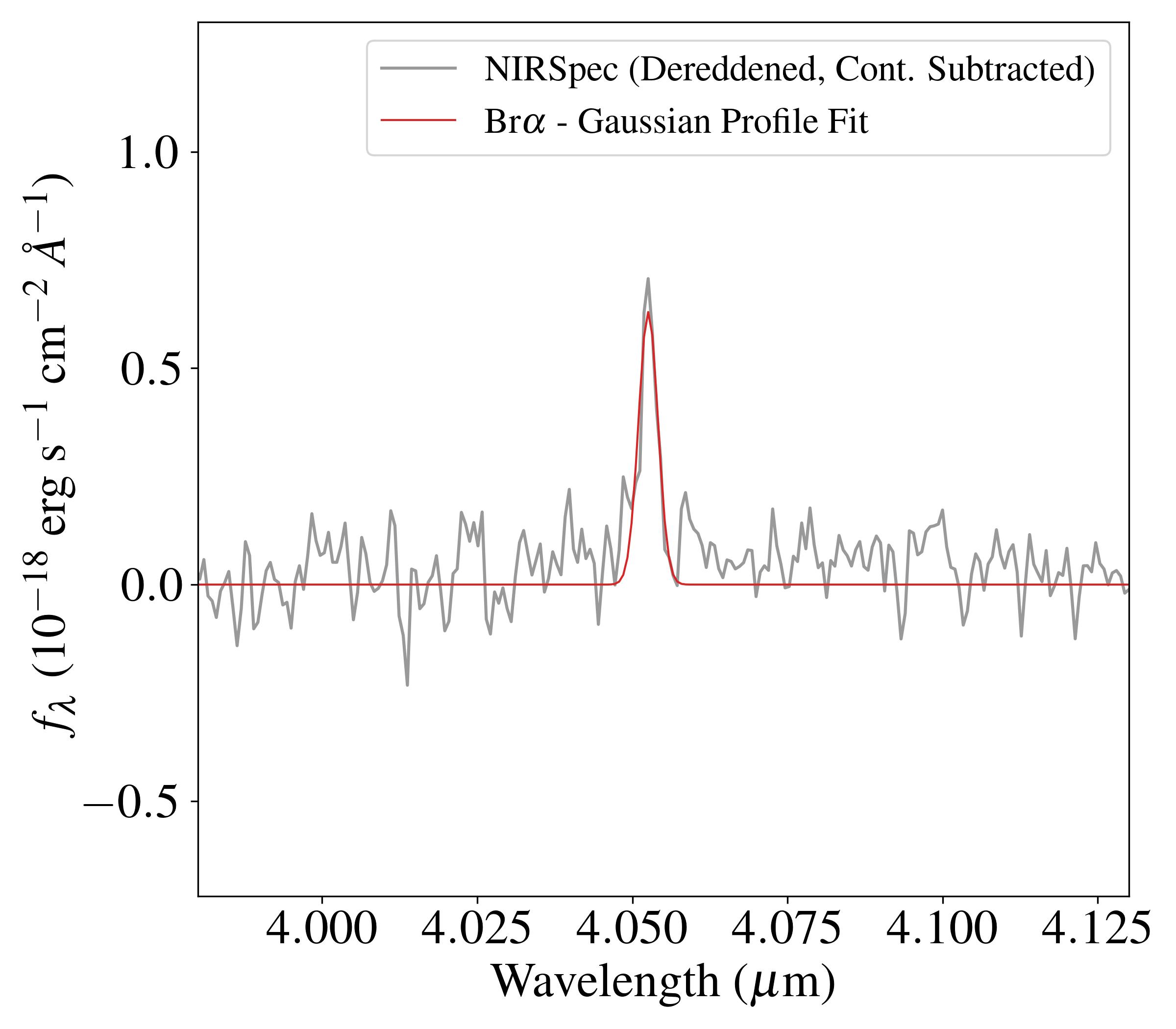}
    \caption{Best-fit \rev{Gaussian} profile model of the Br$\alpha$ emission line detected from \WT~overlaid on the dereddened and continuum-subtracted JWST/NIRSpec spectrum.} 
    \label{fig:Bra}
\end{figure}

The NIRSpec spectrum reveals a $\sim$\rev{10}\,$\sigma$ detection of the Br$\alpha$ hydrogen recombination line from \WT. Previous optical and near-IR spectra of \WT~during its outburst presented by \citet{De2023} notably did not capture any atomic lines in emission. 
Utilizing the \texttt{specutils} Python package, we fit a 
\rev{Gaussian} profile to the Br$\alpha$ line from the derredenned and continuum-subtracted NIRSpec spectrum (Fig.~\ref{fig:Bra}). The properties of the best-fit Br$\alpha$ emission line model are shown in Table~\ref{tab:Bralpha}.
The detection of Br$\alpha$ suggests the presence of hot circumstellar gas, perhaps accreting onto the remnant star.

\begin{deluxetable}{ll}
\tablecaption{\WT~Br$\alpha$ Emission Line Properties}
\tablewidth{0pt}
\tablehead{Parameter & Value }\
\startdata
\rev{Amplitude} & \rev{$(6.30\pm0.49)\times10^{-19}$ erg s$^{-1}$ cm$^{-2}$ $\AA^{-1}$} \\
\rev{Peak Position} & \rev{$4.0526 \pm 0.0001$ $\mu$m} \\
\rev{FWHM} & \rev{$0.0016 \pm 0.0001$ $\mu$m} \\
\rev{Line Flux}  & \rev{$(2.5\pm0.4) \times 10^{-17}$ erg s$^{-1}$ cm$^{-2}$} \\
\enddata
\tablecomments{Derived Br$\alpha$ emission line properties from the best-fit \rev{Gaussian} model. }
\label{tab:Bralpha}
\end{deluxetable}

\begin{figure*}[t]
    \includegraphics[width=0.98\linewidth]{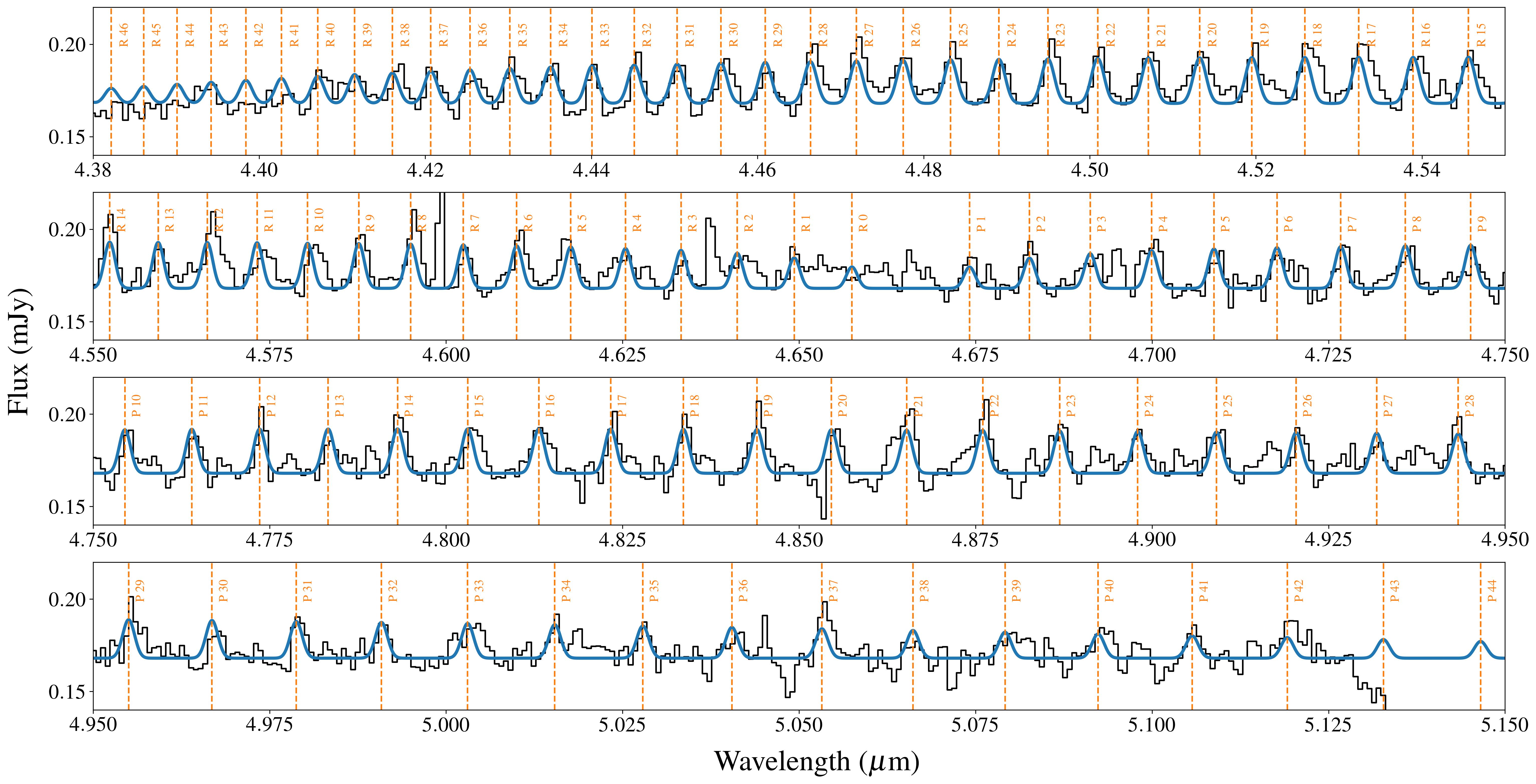} 
    \caption{\rev{Dereddened} NIRSpec G395H/F290LP spectrum of the $^{12}$CO fundamental band emission between 4.38 -- 5.15 $\mu$m from \WT~overlaid with the $^{12}$CO model spectrum from \texttt{spectools\_ir} and the wavelengths of the ``R''- and ``P''-branch transitions. } 
    \label{fig:CO}
\end{figure*}

\begin{figure}[t]
    \includegraphics[width=0.98\linewidth]{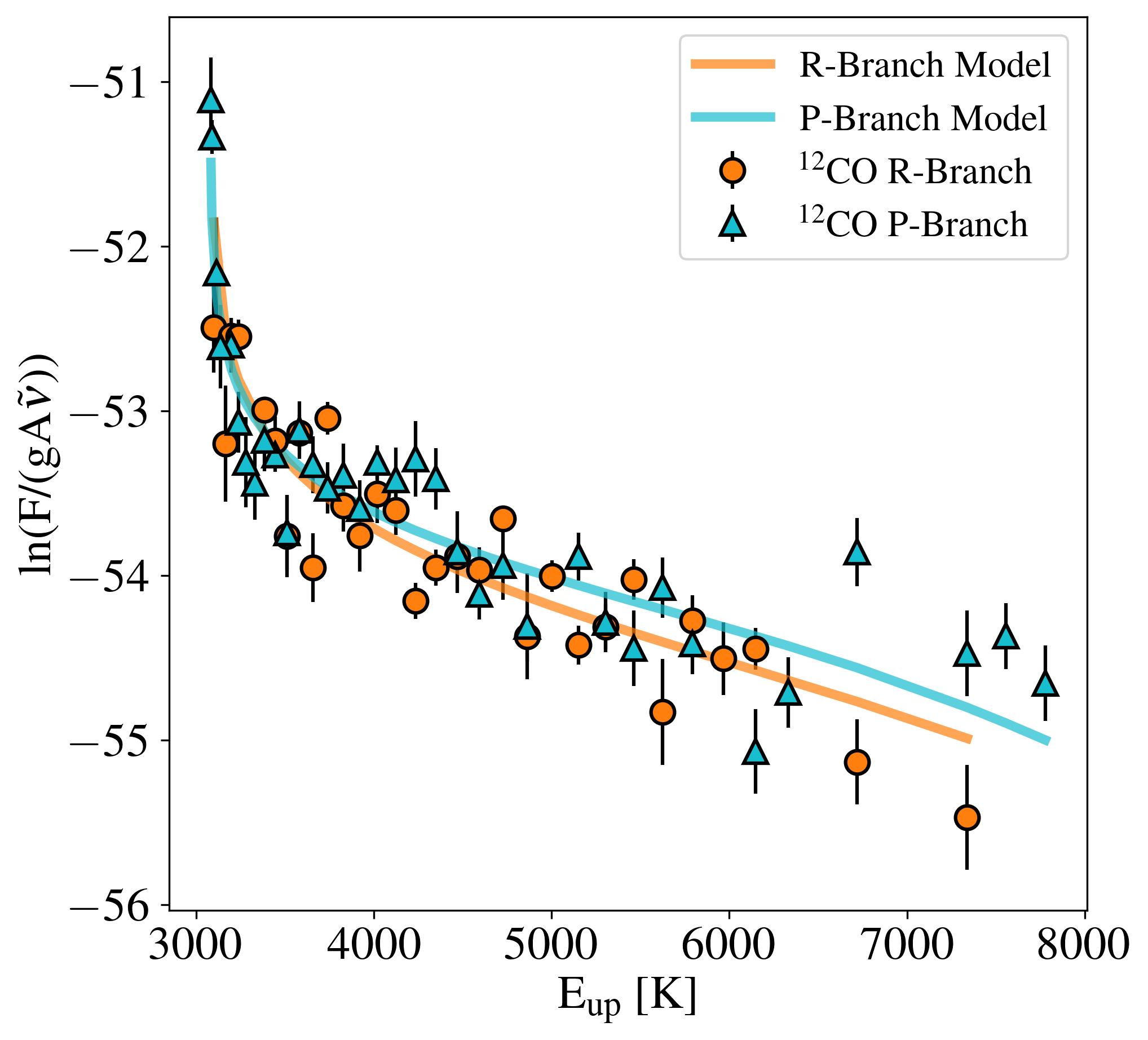} 
    \caption{Rotation diagram (in CGS units) of the measured $^{12}$CO R- and P-branch transitions from the NIRSpec spectrum of \WT~and the model output from \texttt{spectools\_ir}. $F$ is the integrated line flux and E$_\mathrm{up}$ is the upper level energy of the transition.} 
    \label{fig:RD}
\end{figure}

\rev{A broad emission feature at $\sim4.3$ $\mu$m is present in the NIRSpec spectrum which we suggest is associated with \RML{the strongest} emission band of phosphine (PH$_3$; See~\citealt{Sousa-silva2020}). PH$_3$ has previously been detected in the atmospheres of the gas giant planets Jupiter and Saturn \citep{Bregman1975,Larson1977} as well as the circumstellar environment of the carbon-rich asymptotic giant branch (AGB) star IRC +10216 \citep{Agundez2014}. Recent observations with JWST have also indicated the presence of PH$_3$ in the atmosphere of a brown dwarf based on the absorption signature of PH$_3$ between $4.0-4.4$ $\mu$m \citep{Burgasser2024}. \RML{The second strongest PH$_3$ feature is a broad band between 8 - 11 $\mu$m \citep{Sousa-silva2020} is not present in the MIRI LRS spectrum of \WT. However, that PH$_3$ feature would be difficult to detect since the 8 - 11 $\mu$m wavelength range is dominated by thermal dust emission (See Sec.~\ref{sec:830}).} Analysis of the suggested PH$_3$ feature from \WT~is presented in Sec.~\ref{sec:PH3}. Another broad emission feature appears at $\sim4.35$ $\mu$m that is unlikely related to PH$_3$. The identification of the $\sim4.35$ $\mu$m feature is uncertain, but the feature may be blended with $^{12}$CO fundamental band emission. The apparent narrow absorption ``feature'' between the 4.3 and 4.35 $\mu$m features is likely an artifact due to a bad pixel.}

Utilizing \texttt{spectools\_ir}, a suite of tools designed for analysis of molecular astronomical IR spectra, we confirm that the series of emission features around $\sim4.7$ $\mu$m in the NIRSpec spectrum of \WT~originates from $^{12}$CO fundamental band emission. A model of the $^{12}$CO fundamental band emission lines from \texttt{spectools\_ir} is overlaid on the 4.4 - 5.1 $\mu$m spectrum of \WT~in Figure~\ref{fig:CO}. Further analysis and details on the CO modeling is discussed in Sec.~\ref{sec:CO}.
The $^{12}$CO fundamental band emission supports also supports presence of hot gas.




\subsection{$^{12}$CO Emission Modeling}
\label{sec:CO}

We investigated the physical properties of circumstellar gas probed by the $^{12}$CO fundamental band emission of \WT~utilizing the ``slab'' fitting tools \texttt{slabspec} and \texttt{slab\_fitter} in \texttt{spectools\_ir}\footnote{This package is available on The Python Package Index (\url{https://pypi.org/project/spectools-ir/}) and version 1.0.0 is archived on Zenodo \citep{Salyk2022}.}. The distance to \WT~was assumed to be $d = 4$ kpc \citep{De2023}. 
The \texttt{flux\_calculator} tool in \texttt{spectools\_ir} was used to compute CO line fluxes assuming Gaussian line profiles. The \texttt{slab\_fitter} tool then uses the Markov Chain Monte Carlo (MCMC) fitting code \texttt{emcee} \citep{FM2013} to fit a slab model to the CO line fluxes measured from \texttt{flux\_calculator}.

Based on the \texttt{slab\_fitter} model to the CO emission features, we were able to derive the column density of emitting gas, $N_\mathrm{CO}$, radius of the emitting region assuming a circular geometry, $R_\mathrm{C}$, and the excitation temperature, $T_\mathrm{CO}$. We then used \texttt{slabspec} to generate the $^{12}$CO emission model spectrum in Figure~\ref{fig:CO} that was convolved with a $\Delta v = 120$ km s$^{-1}$ FWHM Gaussian to match the approximate resolution of the NIRSpec spectrum.
The model calculations and NIRSpec-detected lines of the $^{12}$CO R- and P-branch transitions from \WT~are shown in the rotation diagram in Figure~\ref{fig:RD}. \RML{The $^{12}$CO emission model spectrum shows a close agreement with the observed features in the 4.4 - 5.1 $\mu$m wavelength range. Apparent emission features that are not reproduced by the model (e.g.~See Fig.~\ref{fig:CO} between the R2 and R3 transitions) may be due to instrumental artifacts but warrant further investigation in a future study.}

The best-fit \texttt{slab\_fitter} model to the $^{12}$CO emission from \WT~provided an excitation temperature of \rev{T$_\mathrm{CO}=1340^{+130}_{-110}$} K with an emitting area of \rev{Log$\left(\frac{\mathrm{\Omega}_\mathrm{CO}} {\mathrm{sr}}\right)= -19.64^{+0.11}_{-0.12}$ (or $9.8^{+2.8}_{-2.4}\times10^{-10}$ arcsec$^{2}$)} and a column density of \rev{Log$\left(\frac{\mathrm{N}_\mathrm{CO}} {\mathrm{cm}^{-2}}\right)=18.02^{+0.21}_{-0.16}$}. Assuming a circular disk geometry for the emitting region of $^{12}$CO, the disk radius is \rev{$R_\mathrm{CO}=15.2^{+2.0}_{-2.0}$ R$_\odot$}. The total $^{12}$CO mass, $M_\mathrm{CO}$, from such a disk can then be estimated as $\mathrm{N}_\mathrm{CO} \times \pi \mathrm{R}_\mathrm{CO}^2 \times m_\mathrm{CO}$, where $m_\mathrm{CO}$($=4.65128\times10^{-23}$ g) is the mass of a $^{12}$CO molecule. We therefore estimate

\begin{equation}
\rev{\mathrm{Log}\left(\frac{M_\mathrm{CO}}{\mathrm{M}_\odot}\right)\approx-13.1^{+0.3}_{-0.3},}
\end{equation}

\noindent
Assuming that the gas traced by the $^{12}$CO emission is fully molecular, approximately exhibits a solar abundance, and that the mass is dominated by molecular hydrogen H$_2$, the total gas mass can be estimated as $\mathrm{M}_\mathrm{CO} \times N(H_2/^{12} CO) \times m_{\mathrm{H}_2} / m_\mathrm{CO}$, where $N(H_2/^{12} CO)$ is assumed to be $10^4$ and $m_{\mathrm{H}_2} / m_\mathrm{CO}$ is $\approx0.071$. We estimate a total gas mass of


\begin{equation}
\rev{\mathrm{Log}\left(\frac{M_\mathrm{gas}}{\mathrm{M}_\odot}\right)\approx-10.2^{+0.3}_{-0.3}.}
\end{equation}


Detection of the $^{12}$CO fundamental band ($\nu=1-0$) in emission is typically seen from embedded young stellar objects (YSOs; \citealt{Pontoppidan2003,Herczeg2011}), disks around Herbig Ae/Be stars \citep{Blake2004}, and T Tauri disks \citep{Salyk2011}. However, $^{12}$CO fundamental band emission has notably been detected around V4332 Sgr \citep{Banerjee2004}, a red nova that has been compared to \WT~given its low luminosity \citep{De2023}. 

\WT~is unlikely associated with a YSO outburst or Herbig Ae/Be star due to the lack of atomic emission lines in the optical and near-IR spectrum taken during outburst \citep{De2023}. The CO excitation temperature we measure for \WT~is also much hotter than that observed from T Tauri disks \citep{Salyk2011}. The JWST spectra also do not exhibit any prominent ice absorption features, which are common for embedded YSOs. Additionally, the lack of polycyclic aromatic hydrocarbon (PAH) features between 6-12 $\mu$m in the LRS spectrum of \WT~likely rules out a Herbig Ae/Be origin given that they commonly exhibit PAH features \citep{Keller2008}.
The JWST spectroscopic observations are therefore consistent with the claim that \WT~arose from a planetary engulfment event.

\begin{figure*}[t]
    \includegraphics[width=0.98\linewidth]{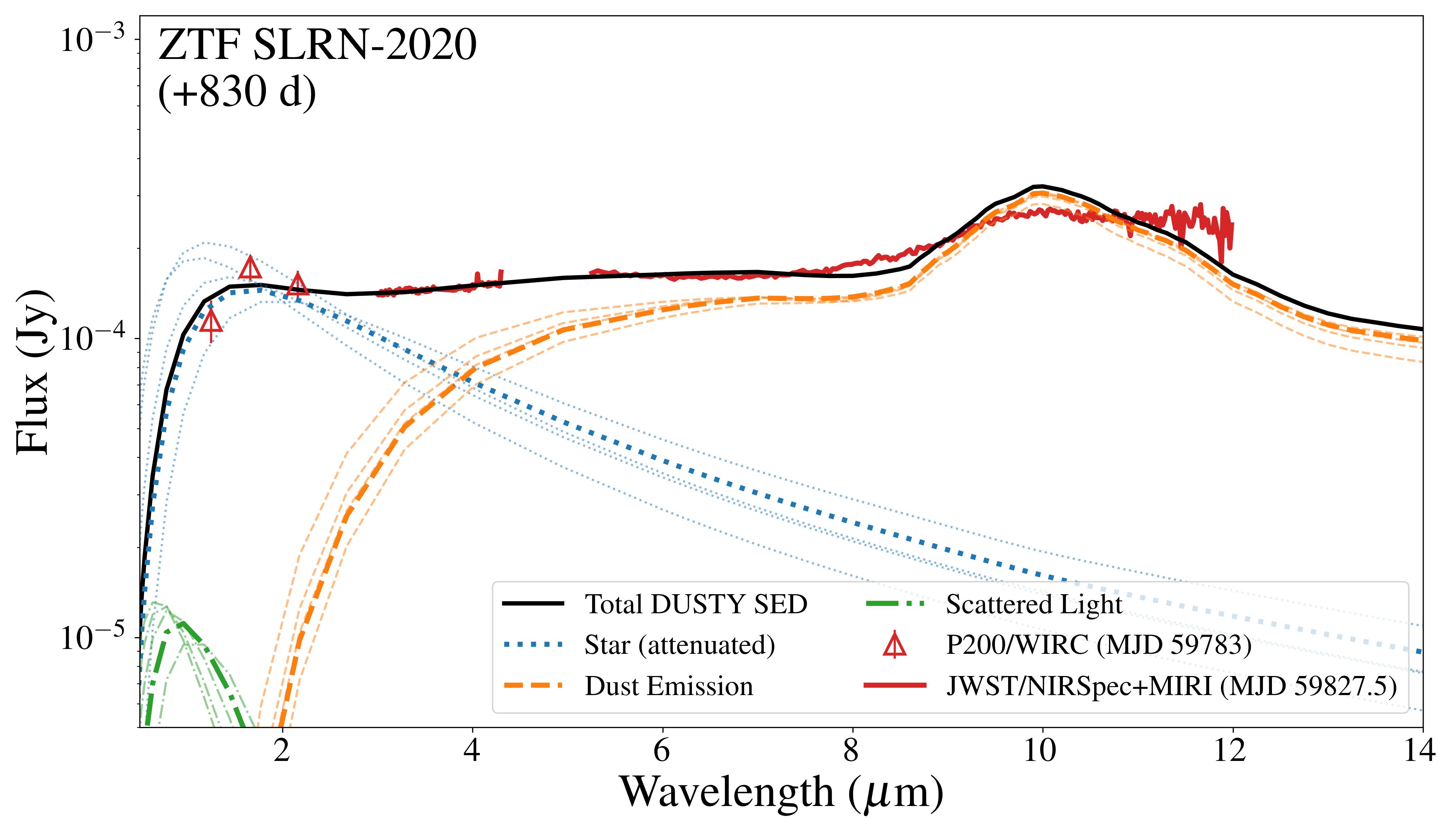}
    \caption{Best-fit \texttt{DUSTY} models of the \WT~SED from JWST and P200/WIRC data (+830 d epoch) using five different effective temperatures for the heating source (i.e.~remnant star): T$_*=3000$, 3500, 4000, 4500, and 5000 K. The T$_*=3500$ K model (bold lines) provided the best fit to the P200/WIRC near-IR photometry. The solid black line shows the total T$_*=3500$ K \texttt{DUSTY} SED model composed of the attenuated stellar emission (blue, dotted), dust emission (orange, dashed), and dust-scattered starlight (green, dot-dashed).} 
    \label{fig:DUSTY}
\end{figure*}

\begin{deluxetable}{p{6cm}l}
\tablecaption{Model Results Summary}
\tablewidth{0pt}
\tablehead{Parameter & Value}\
\startdata
\hline
\texttt{spectools\_ir} Best-fit $^{12}$CO Model\\ (+830 d, JWST)\\
T$_\mathrm{CO}$ & \rev{$1340^{+130}_{-110}$ K} \\
Log$\left(\frac{\mathrm{\Omega}_\mathrm{CO}} {\mathrm{sr}}\right)$ & \rev{$-19.64^{+0.11}_{-0.12}$} \\
R$_\mathrm{CO}$ & \rev{$15.2^{+2.0}_{-2.0}$ R$_\odot$} \\
Log$\left(\frac{\mathrm{N}_\mathrm{CO}} {\mathrm{cm}^{-2}}\right)$ & \rev{$18.02^{+0.21}_{-0.16}$} \\
Log$\left(\frac{M_\mathrm{CO}}{\mathrm{M}_\odot}\right)$ & \rev{$-13.1^{+0.3}_{-0.3}$} \\
Log$\left(\frac{M_\mathrm{gas}}{\mathrm{M}_\odot}\right)$ & \rev{$-10.2^{+0.3}_{-0.3}$} \\
\hline
\texttt{DUSTY} Best-fit SED Model\\ (+830 d, JWST + P200)\\
L$_*$ & \rev{$0.29^{+0.03}_{-0.06}$ L$_\odot$} \\
T$_\mathrm{d}$ & \rev{$720^{+80}_{-50}$ K} \\
$\tau_V$ & \rev{$0.7^{+0.3}_{-0.1}$} \\
T$_*$ & 3500 K \\
R$_\mathrm{in}$ & \rev{$50.5^{+9.5}_{-11.0}$ R$_\odot$} \\
Log(M$_\mathrm{d}$) & \rev{$-10.61^{+0.08}_{-0.16}$ M$_\odot$} \\
\hline
\texttt{DUSTY} Best-fit SED Model \\(+320 d, ZTF + P200 + NEOWISE)\\
L$_*$ & $13.06^{+5.61}_{-10.76}$ L$_\odot$ \\
T$_\mathrm{d}$ & $280^{+450}_{-20}$ K \\
$\tau_V$ & $4.2^{+0.4}_{-3.6}$ \\
T$_*$ & $5800^{+800}_{-2800}$ K \\
R$_\mathrm{in}$ & $4730^{+1600}_{-4420}$ R$_\odot$ \\
Log$\left(\frac{M_\mathrm{d}}{\mathrm{M}_\odot}\right)$ & $-5.89^{+0.29}_{-3.21}$ \\
\enddata
\tablecomments{Best-fit model parameters from IR observations of \WT. The \texttt{spectools\_ir} results were fit to the $^{12}$CO fundamental band emission revealed by the JWST/NIRSpec spectrum taken +830 d from the emission peak. \texttt{DUSTY} SED models were fit to the dereddened, +830 d observations from NIRSpec, MIRI, and P200/WIRC. \texttt{DUSTY} SED models were also fit to dereddened archival +320 d observations taken by P48+ZTF, P200/WIRC, and NEOWISE +320 d from the emission peak. The interstellar extinction and distance of \WT~were assumed to be $A_V=3.6$ and $d=4$ kpc, respectively \citep{De2023}. }  
\label{tab:Results}
\end{deluxetable}

\subsection{Circumstellar Dust Emission Modeling}
\label{sec:DUSTY}



\subsubsection{SED Modeling of +830 d Epoch}
\label{sec:830}

In order to constrain the luminosity and effective temperature of the remnant star and investigate the circumstellar dust around \WT, we used the \texttt{DUSTY} radiative transfer code \citep{DUSTY} to model its full 1 - 12 $\mu$m SED. We fit \texttt{DUSTY} SED models to the dereddened data that includes the P200/WIRC JHK$_s$ photometry, smoothed NIRSpec 3--5 $\mu$m spectra with the $\sim4.7$ $\mu$m $^{12}$CO fundamental band emission removed, and the 5--12 $\mu$m LRS spectrum. Note that the $^{12}$CO emission was removed from the fitting since \texttt{DUSTY} only models the emission from the heating source, scattered light, and thermal emission from circumstellar dust. 

Using the \texttt{SPHERE} geometry, the input parameters related to the circumstellar dust were the dust composition, grain size distribution, sublimation temperature ($T_\mathrm{Sub}$), dust temperature at the inner boundary ($T_\mathrm{d}$), the shell thickness factor ($Y$)\footnote{i.e. the outer radius of the dust shell is R$_\mathrm{in}\times Y$}, the dust density power-law of the shell ($\alpha$), and the optical depth $\tau_V$. The central radiative dust-heating source is assumed to be a blackbody with an effective temperature $T_*$, which is a free input parameter for the modeling. \texttt{DUSTY} outputs an SED model that includes the attenuated and non-attenuated heating-source spectrum, dust-scattered light, and thermal dust emission. \texttt{DUSTY} also calculates an inner dust shell radius (R$_\mathrm{in}$) that assumes a fixed heating-source luminosity of $10^4$ L$_\odot$ with an effective temperature T$_*$. The output SED must then be normalized to the 1 - 12 $\mu$m SED of the \WT~to derive the heating-source luminosity and the inner shell radius. We therefore fit a multiplicative factor to scale the output SED to the \WT~SED. Adopting a distance of $d = 4$ kpc, the total stellar luminosity L$_*$ can then be derived by integrating over the non-attenuated heating source spectrum scaled to the data. We utilized a reduced-$\chi^2$ analysis to characterize the goodness-of-fit of the \texttt{DUSTY} SED model to the \WT~SED.

A standard \citet{Mathis1977} (MRN) distribution was adopted for the grain sizes that assumes minimum and maximum grain sizes of 0.005 $\mu$m and 0.25 $\mu$m, respectively, and a grain-size power-law index of -3.5. Given the uncertainties in the grain composition, a 50/50 silicate/amorphous-carbon composition was adopted using the \texttt{Sil-Ow} and \texttt{amC-Hn} optical properties provided by \texttt{DUSTY} \citep{Ossenkopf1992,Hanner1988}. The dust sublimation temperature was fixed at value of $T_\mathrm{Sub}=1500$ K. Lastly, a $\propto r^{-2}$ radial dust-density profile and a shell thickness factor of $Y=5$ was adopted for the spherically symmetric circumstellar dust. 

We note that the \texttt{DUSTY} SED analysis conducted by \citet{De2023} on \WT~also adopted the same grain size distribution, radial density profile, and shell thickness\rev{, but with a pure silicate dust composition. A pure silicate dust composition was attempted in our modeling but led to unsatisfactory fitting results. We emphasize the uncertainty on the grain composition given that the shape of the 9.7 $\mu$m silicate feature largely influenced from the interstellar extinction correction (Fig.~\ref{fig:SED}) }

We performed the SED-fitting by conducting a grid search across the inner boundary dust temperature $T_\mathrm{d}$ and the optical depth of the circumstellar dust shell $\tau_V$. The values for $T_\mathrm{d}$ ranged from 500--1500 K with a step size of 25 K. The values for $\tau_V$ ranged from 0.1--5.0 with a step size of 0.1. Due to the limited coverage of the near-IR photometry probing direct emission from the heating source, the effective temperature of the heating source $T_*$ is difficult to constrain with \texttt{DUSTY} models. We therefore performed the $T_\mathrm{d}$-$\tau_V$ grid search described above using five different effective temperatures for the heating source (i.e. remnant star): T$_*=$ 3000, 3500, 4000, 4500, and 5000 K. The results of the best-fit DUSTY SEDs for all five effective temperatures are shown in Figure~\ref{fig:DUSTY}, where the T$_*=3500$ K model exhibited the best fit to the P200/WIRC near-IR photometry. We therefore favor the best-fit results from the T$_*=3500$ K model, which are presented in Table~\ref{tab:Results}.


\begin{figure*}[t]
    \includegraphics[width=0.98\linewidth]{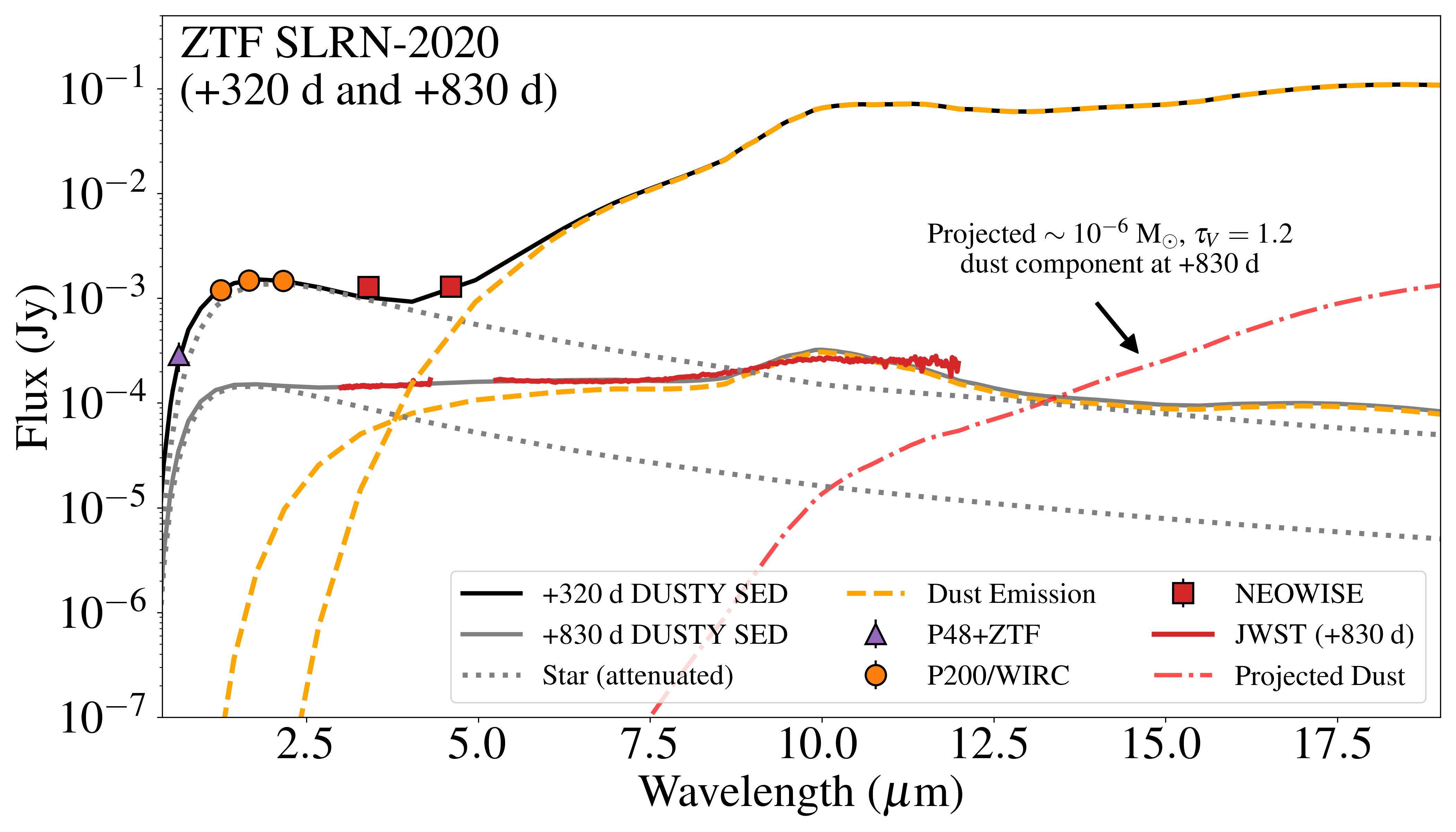}
    \caption{Best-fit \texttt{DUSTY} model of the \WT~SED from archival P48+ZTF, P200/WIRC, and NEOWISE data taken +320 d after the outburst peak \citep{De2023} overlaid with the JWST (+830 d) SED and best-fit, T$_*=3500$ K model from Figure~\ref{fig:DUSTY}. The best-fit \rev{+320 d} parameters of the model are provided in Tab.~\ref{tab:Results}. The \rev{dot-dashed red} line shows the estimated SED of the +320 d model projected to +830 d, which assumes a dust mass of $\sim10^{-6}$ M$_\odot$ and an optical depth of $\tau_V=1.2$ (See Sec.~\ref{sec:constraints}). This projected dust component could be undetected in the +830 d JWST observations. \rev{For clarity, the scattered-light emission component from the \texttt{DUSTY} models are not shown but are included in the total \texttt{DUSTY} SED models (solid black and grey lines).}} 
    \label{fig:320sed}
\end{figure*}

The luminosity of the remnant star from the best-fit \texttt{DUSTY} SED models of \WT~is \rev{L$_*=0.29^{+0.03}_{-0.06}$ L$_\odot$}.
If this derived luminosity is representative of the star's baseline output, it suggests
 that \WT~hosts a $\sim0.7$ M$_\odot$ K-type star\footnote{Assuming a mass-luminosity relation of $\left(\frac{L}{L_\odot}\right)\approx\left(\frac{M}{M_\odot}\right)^4$.}. The low mass derived by the \texttt{DUSTY} analysis of the +830 d data is in rough agreement with the lower mass estimate range inferred from the IR progenitor brightness and colors by \citet{De2023} ($0.8-1.5$ M$_\odot$), who used archival data from the United Kingdom Infrared Telescope (UKIRT) Galactic plane survey \citep{Lawrence2007}. \citet{De2023} also noted that the stellar masses were not well constrained due to the photometric errors of the near-IR UKIRT progenitor photometry: $\sigma_H$ = 0.25 mag and $\sigma_K=0.18$ mag. 
 
 Intriguingly, a K-type star should still be on the main-sequence, which suggests that radial expansion due to stellar evolution may not have been the mechanism that triggered the planetary engulfment event. However, it is possible that a cool and optically thick circumstellar dust component \citep{De2023} undetected by JWST may be obscuring a fraction of the stellar luminosity. We investigate this in more detail in Sec.~\ref{sec:obscured}. 

We derived the circumstellar dust mass of \WT~from the best-fit \texttt{DUSTY} SED modeling parameters (Tab.~\ref{tab:Results}) and following the approach outlined in Sec.~\ref{sec:appendix} and Eq.~\ref{eq:mass3}:


\begin{equation}
    \rev{\mathrm{Log}\left(\frac{M_\mathrm{d}}{\mathrm{M}_\odot}\right) \approx -10.61^{+0.08}_{-0.16}.}
\label{eq:830mass}
\end{equation}

\noindent
In the dust mass calculations, a value of $1.123\times10^4$ cm$^2$ g$^{-1}$ is adopted for the dust opacity\footnote{the dust absorption cross section weighted by dust mass} ($\kappa_V^{d}$), which is consistent with the \citet{WD2001} Milky Way grain-size distribution for $R_V=3.1$ with dust optical properties from \citet{Draine2003}\footnote{\url{https://www.astro.princeton.edu/~draine/dust/extcurvs/kext_albedo_WD_MW_3.1_60_D03.all}}. 

Assuming a gas-to-dust mass ratio of 100, we can estimate the total mass of the circumstellar material around \WT:

\begin{equation}
    \rev{\mathrm{Log}\left(\frac{M_\mathrm{CSM}}{\mathrm{M}_\odot}\right) \approx -8.61^{+0.08}_{-0.16}.}
\label{eq:warm_csm}
\end{equation}

\noindent
The circumstellar mass is notably several orders of magnitude less than the $\sim10^{-6}$ M$_\odot$ of material estimated by \citet{De2023} from their \texttt{DUSTY} SED modeling of the +320 d SED of \WT. The dust temperature and inner radius from our best-fit \texttt{DUSTY} model of the +830 d SED ($T_\mathrm{d}\approx720$ K and R$_\mathrm{in}\approx50$ R$_\odot$; Tab.~\ref{tab:Results}) is hotter and smaller, respectively, than the values derived by \citet{De2023} from the +320 d SED, where $T_\mathrm{d}=415$ K and R$_\mathrm{in}=1140$ R$_\odot$. 
The less massive, hotter, and more close-in circumstellar material derived from the +830 d epoch compared to that of the +320 d epoch may therefore be a distinct circumstellar component surrounding the remnant star that was not revealed by the +320 d SED \citep{De2023}. 


\subsection{Revisiting the +320 d SED Modeling}
\label{sec:320}
In order to investigate the evolving properties of \WT~between the +320 d epoch reported by \citet{De2023} and the +830 d epoch from this work, we revisit the +320 d SED using a consistent \texttt{DUSTY} modeling approach that we performed for the +830 d epoch. Utilizing the photometry from $r$-band through W2 (i.e.~$\lambda = 0.64 - 4.6$ $\mu$m) reported by \citet{De2023} from the +320 d epoch, we conducted a grid search across the inner boundary dust temperature $T_\mathrm{d}$, the optical depth of the circumstellar dust shell $\tau_V$, and the effective temperature of the heating source $T_*$. Note that in this analysis of the +320 d SED, the effective temperature was included as a grid-search parameter due to the better wavelength coverage of the heating source compared to that of the +830 d observations. We also adopt the same grain size and composition parameters as the +830 d model: an MRN distribution with minimum and maximum grain sizes of 0.005 $\mu$m and 0.25 $\mu$m, respectively, a grain-size power-law index of -3.5, and a 50/50 silicate/amorphous-carbon composition.

In the +320 d \texttt{DUSTY} grid search, the values for $T_\mathrm{d}$ ranged from 100--1000 K with a step size of 25 K, the values for $\tau_V$ ranged from 0.2--15 with a step size of 0.4, and the values for $T_*$ ranged from 2600--7000 K with a step size of 200 K. Figure~\ref{fig:320sed} shows the resulting best-fit model for the +320 d SED and the +830 d SED model for comparison. The best-fit SED model parameters are provided in Table~\ref{tab:Results}. 

The luminosity of \WT~was notably a factor of $\gtrsim10$ greater at +320 d, when \WT~was still in outburst, than at +830 d. 
The rapid decrease in luminosity deviates from the predicted $L_*\propto t^{-4/5}$ power-law decay for the gravitational contraction of an inflated host star envelope following the engulfment \citep{Tylenda2005, De2023}.

The dust mass of the \WT~ejecta probed by the +320 d SED, Log$\left(\frac{M_\mathrm{d}}{\mathrm{M}_\odot}\right)=-5.89^{+0.29}_{-3.21}$, implies a total ejecta mass of 

\begin{equation}
\mathrm{Log}\left(\frac{M_\mathrm{ej}}{\mathrm{M}_\odot}\right)=-3.89^{+0.29}_{-3.21}
\label{eq:ejmass}
\end{equation}
\smallskip

\noindent
assuming a gas-to-dust mass ratio of 100. 
Although the best-fit model parameters for the +320 d exhibits large uncertainties due to the limited mid-IR wavelength coverage out to 4.6 $\mu$m, the results suggest that there is a more extended, cooler, and more massive circumstellar component and a closer-in, hotter, and less massive component. 
We note that the implied total ejecta mass of $\sim10^{-4}$ M$_\odot$ ($\approx0.1$ M$_\mathrm{Jup}$) is consistent with the inferred ejecta mass based on the luminosity and duration of the \WT~outburst \citep{Matsumoto2022,De2023}.


\begin{figure}[t]
    \includegraphics[width=0.98\linewidth]{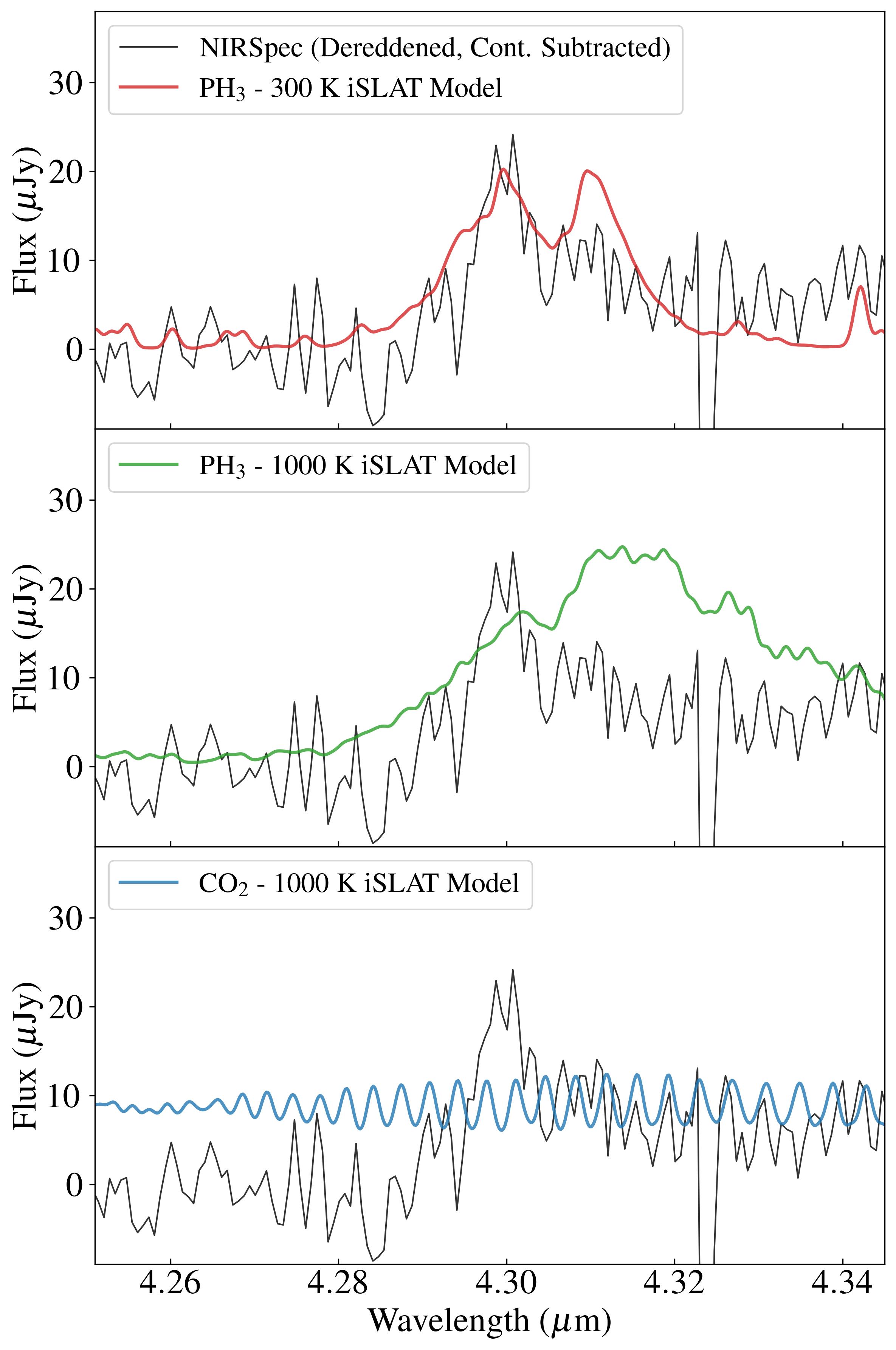}
    \caption{\rev{Synthetic iSLAT spectra of (\textit{Top}) PH$_3$ at 300 K, (\textit{Middle}) PH$_3$ at 1000 K , and (\textit{Bottom}) CO$_2$ at 1000 K overlaid on the dereddened and continuum-subtracted JWST/NIRSpec spectrum of \WT. The 300 K PH$_3$ model presents the closest agreement with the NIRSpec spectrum.}} 
    \label{fig:PH3}
\end{figure}


\subsection{\rev{Possible detection of Phosphine (PH$_3$)}}
\label{sec:PH3}


\rev{Based on an analysis with the interactive Spectral-Line Analysis Tool (iSLAT, \citealt{Jellison2024}), we attributed the broad emission feature revealed by NIRSpec at $\sim4.3$ $\mu$m to PH$_3$\footnote{\rev{iSLAT model spectra were convolved with a Gaussian with a FWHM of 120 km s$^{-1}$ to match the resolving power of the NIRSpec observations.}}. Synthetic spectra of PH$_3$ at temperatures consistent with warm (1000 K) and cool (300 K) \RML{models} were generated from iSLAT using a simple slab model and molecular data from HITRAN \citep{HITRAN}.  The synthetic PH$_3$ spectra are shown in Figure~\ref{fig:PH3} overlaid on the dereddened and continuum-subtracted $\sim4.3$ $\mu$m feature. Since CO$_2$ has a prominent band that overlaps with the $\sim4.3$ $\mu$m PH$_3$ band, we also generated a synthetic spectrum of CO$_2$ to rule it out as the origin of the feature (Fig.~\ref{fig:PH3}). Neither the 1000 K PH$_3$ model nor the CO$_2$ model resemble the $\sim4.3$ $\mu$m feature; however, the 300 K PH$_3$ model shows a general agreement with the feature.}

\rev{The 300 K PH$_3$ iSLAT model assumed a radius of 22 au, consistent with the inner radius of the best-fit \texttt{DUSTY} model of the ejecta dust (i.e.~the dust component revealed at +320 d; Tab.~\ref{tab:Results}). A column density of $\mathrm{N}_{\mathrm{PH}_3}\sim2\times10^{15}$ cm$^{-2}$ was derived by approximately scaling the amplitude of the synthetic feature to that of the observed feature. If the emitting PH$_3$ is indeed located in the ejecta, we can estimate the PH$_3$ mass and the abundance of PH$_3$ relative to H$_2$ assuming that the total gas mass of the ejecta (M$_\mathrm{ej}\sim10^{-4}$ M$_\odot$; See Eq.~\ref{eq:ejmass}) is dominated by H$_2$. The approximate PH$_3$ mass is}

\begin{equation}
\rev{\mathrm{N}_{\mathrm{PH}_3}\times \pi \mathrm{R}_{\mathrm{PH}_3}^2\times m_{\mathrm{PH}_3}\sim2\times10^{-11}\,\mathrm{M}_\odot}
\end{equation}

\noindent
\rev{and the abundance of PH$_3$ relative to H$_2$ is approximately}

\begin{equation}
\rev{\frac{\mathrm{N}_{\mathrm{PH}_3} \times \pi \mathrm{R}_{\mathrm{PH}_3}^2}{\mathrm{M}_\mathrm{ej} / \mathrm{m}_{\mathrm{H}_2}}\sim 10^{-8}}
\end{equation}

\noindent
\rev{where m$_{\mathrm{H}_2}$ is the mass of an H$_2$ molecule, m$_{\mathrm{PH}_3}$ is the mass of a PH$_3$ molecule, and R$_{\mathrm{PH}_3}$ ($=22$ au) is the assumed radius of the PH$_3$ emitting region. Interestingly, the abundance of PH$_3$ relative to H$_2$ from \WT~is consistent with that of the circumstellar material around the AGB star IRC +10216 \citep{Agundez2014}. The molecular outflows from \WT~and stellar mergers notably exhibit similar molecular features as the cool envelopes of red, evolved stars \citep{Kaminski2018,De2023}, which suggests similar conditions for PH$_3$ formation. The consistent relative PH$_3$ abundances provides further evidence that the $\sim4.3$ $\mu$m feature from \WT~is indeed associated with PH$_3$. }

\section{Discussion}
\label{sec:discussion}

\subsection{Investigating the Remnant Host Star and its Circumstellar Environment}
There are two prevailing and related questions arising from the newly obtained +830 d observations from JWST and the re-analysis of the +320 d SED from \citet{De2023}:

\begin{itemize}
\itemsep0em 
    \item Does \WT~host a $\sim0.7$ M$_\odot$ K-type star that should not have evolved off the main sequence yet or is the stellar luminosity obscured by cooler circumstellar dust?
    \item Could $\sim10^{-6}$ M$_\odot$ of dust ($\approx10^{-4}\, \mathrm{M}_\odot$ of total gas+dust mass) produced by \WT~have gone undetected by the JWST observations?
\end{itemize}

\noindent
These questions are related because the high dust mass component from the +320 d SED model, which is notably absent from the +830 d SED model, may be responsible for blocking a significant fraction of the remnant star luminosity.
In this section, we address these two questions and discuss the implications of our results.  

\subsubsection{Is a More Luminous Remnant Star~Obscured by Cool Circumstellar Dust?}
\label{sec:obscured}


We assess if \WT~might host a more luminous star that could be evolving off the main sequence (e.g.~L$_*\gtrsim1$ L$_\odot$) when accounting for additional obscuration from the $\sim10^{-6}$ M$_\odot$ dust component identified in the +320 d SED but undetected at +830 d (See Tab.~\ref{tab:Results}).  Given the measured luminosity of \rev{L$_*=0.29$ L$_\odot$} from the +830 d SED model, $\gtrsim70\%$ of a $\gtrsim1$ L$_\odot$ source would have to be obscured by the undetected dust component. 
In order to estimate the level of obscuration, we project the +320 d dust optical depth ($\tau^\mathrm{+320d}_V=4.2$) out to +830 d assuming a constant dust expansion velocity of 

\begin{equation}
v_\mathrm{exp}\approx R_\mathrm{in}^\mathrm{+320 d} / \mathrm{320 \, d} \approx 120\, \mathrm{km} \, \mathrm{s}^{-1}
\end{equation}
\noindent
Based on Equations~\ref{eq:tau} \&~\ref{eq:density}, the optical depth can be estimated as follows:

\begin{equation}
    \tau_V^\mathrm{+830d}\approx\tau_V^\mathrm{+320d}\frac{Y^\mathrm{+320d} R_\mathrm{in}^\mathrm{+320d}}{(R_\mathrm{in}^\mathrm{+320d}+\Delta t\, v_\mathrm{exp})(Y^\mathrm{+320d} R_\mathrm{in}^\mathrm{+320d}+\Delta t\, v_\mathrm{exp})},
\end{equation}

\noindent
where $Y^\mathrm{+320d}$ is the shell thickness factor in the +320 d \texttt{DUSTY} models that was assumed to be 5, $\Delta t$ is the time elapsed between the +320 d and +830 d observations (i.e.~$\Delta t = 510$ d), and $R_\mathrm{in}^\mathrm{+320d}$ is the inner radius of $\sim10^{-6}$ M$_\odot$ dust component in the +320 d SED model (Tab.~\ref{tab:Results}). We therefore estimate that the optical depth of the $\sim10^{-6}$ M$_\odot$ component at +830 d would be

\begin{equation}
    \tau_V^\mathrm{+830d}\sim1.2.
\end{equation}

\noindent
Adopting a \citet{WD2001} extinction curve with $\tau_V=1.2$ for the circumstellar dust, we calculate that the dust component would obscure $\approx32-46\%$ of the stellar luminosity for a source with an effective temperature ranging from $3500 - 5000$ K. 
Based on the best-fit \rev{$0.29$} L$_\odot$ luminosity from the +830 d model (Tab.~\ref{tab:Results}), the dust-corrected luminosity of the remnant star would be L$_*^\mathrm{cor}\approx0.4$ L$_\odot$ for an effective temperature of T$_*=3500$ K and L$_*^\mathrm{cor}\approx0.5$ L$_\odot$ for an effective temperature of T$_*=5000$ K. 

The stellar luminosities that have been corrected for possible obscuration from a 10$^{-6}$ M$_\odot$, $\tau_V=1.2$ circumstellar dust shell are notably still consistent with a K-type star with a mass $\lesssim 0.85$ M$_\odot$. Note that these dust-corrected luminosities also agree with the lower-mass range estimated by the IR progenitor brightness and color analysis by \citet{De2023}: $0.8-1.5$ M$_\odot$. These results demonstrate even if the \WT~host star is obscured by an undetected $\sim10^{-6}$ M$_\odot$ of dust at +830 d, it is unlikely the host is massive enough to have left the main sequence. We suggest that the \WT~planetary engulfment event was not triggered by radial expansion from stellar evolution.



\subsubsection{Observational Constraints on a $\sim10^{-6}$ M$_\odot$ Circumstellar Dust Component at +830 d}
\label{sec:constraints}


Based on our \texttt{DUSTY} SED analysis (Sec.~\ref{sec:DUSTY}) and JWST observations, we can estimate constraints on the physical parameters of the \WT~ejecta. Specifically, we address if the $\sim10^{-6}$ M$_\odot$ dust component from the \WT~ejecta identified in the +320 d analysis could have been undetected in the late-time, +830 d JWST observations. 

We utilize \texttt{DUSTY} to model the predicted emission from a $\sim10^{-6}$ M$_\odot$, $\tau_V=1.2$ dust component at +830 d and then compare against the +830 d JWST observations.
Instead of adopting the best-fit +830 d heating source parameters (Tab.~\ref{tab:Obs}), which may be obscured by undetected circumstellar dust, we adopt the dust-corrected star model with L$_*=0.5$ L$_\odot$ and T$_*=5000$ K from Sec.~\ref{sec:obscured}. 
We note that an effective temperature of 5000 K is also inferred for the \WT~host star by \citet{De2023}. The projected shell thickness factor for the $\sim10^{-6}$ M$_\odot$ component can be estimated as $Y^\mathrm{+830d}=\frac{Y^\mathrm{+320d} R_\mathrm{in}^\mathrm{+320d}+\Delta t\, v_\mathrm{exp}}{R_\mathrm{in}^\mathrm{+320d}+\Delta t\, v_\mathrm{exp}}\approx2.5$.

The dust temperature at the inner radius of the $\sim10^{-6}$ M$_\odot$ component at +830 d must then be estimated as the final input needed for the \texttt{DUSTY} model. The temperature of this dust component can be estimated at the time of the +830 d observations assuming it is in radiative equilibrium with the radiation field from the central remnant star. Following the equilibrium temperature description in Sec.~\ref{sec:teq} and adopting a dust emissivity power-law index of $\beta=1$, the projected dust temperature of the $\sim10^{-6}$ M$_\odot$ dust component at +830 can be estimated as follows: 

\begin{equation}
\begin{aligned}
T_\mathrm{d}^\mathrm{+830d} \approx T_\mathrm{d}^\mathrm{+320d} \left(\frac{\left<Q_\mathrm{abs}\right>_\mathrm{*}^\mathrm{+830d}}{\left<Q_\mathrm{abs}\right>_\mathrm{*}^\mathrm{+320d}}\right)^{1/5} \\
\left(\frac{L_*^\mathrm{+830d}}{L_*^\mathrm{+320d}}\right)^{1/5}\left(\frac{r^\mathrm{+830d}}{r^\mathrm{+320d}}\right)^{-2/5},
\end{aligned}
\label{eq:teq830}
\end{equation}

\noindent
where $\left<Q_\mathrm{abs}\right>_\mathrm{*}$ is the spectrum-averaged absorption cross section, L$_*$ is the central heating source luminosity, and $r$ is the separation distance between dust and the heating source. The radius of the $\sim10^{-6}$ M$_\odot$ dust component at +830 d can as estimated as $r^\mathrm{+830d} \approx R_\mathrm{in}^\mathrm{+320d}+\Delta t\, v_\mathrm{exp} \approx 12330$ R$_\odot$, and the heating source luminosity is assumed to be $L_*^\mathrm{+830d}=0.5$ L$_\odot$. The values for $\left<Q_\mathrm{abs}\right>_\mathrm{*}$ of the heating source at +320 d and +830 d can be estimated using Eq.~\ref{eq:Qstar} for effective temperatures T$_*^\mathrm{+320d}=5800$ K (Tab.~\ref{tab:Results}) and T$_*^\mathrm{+830d}=5000$ K. Since the effective temperatures are not too different, we find that $\left(\frac{\left<Q_\mathrm{abs}\right>_\mathrm{*}^\mathrm{+830d}}{\left<Q_\mathrm{abs}\right>_\mathrm{*}^\mathrm{+320d}}\right)^{1/5}\approx1.0$ for silicate and carbonaceous dust grains. 
Using the above parameters and the best-fit model values for $T_\mathrm{d}^\mathrm{+320d}=280$ K, $L_*^\mathrm{+320d}=13$ L$_\odot$, and $r^\mathrm{+320d}=4730$ R$_\odot$, we estimate a projected inner-radius dust temperature of 

\begin{equation}
    T_\mathrm{d}^\mathrm{+830d} \approx 100 \, \mathrm{K}
\end{equation}

\noindent
for the $\sim10^{-6}$ M$_\odot$, $\tau_V=1.2$ dust component at +830 d.

With the input parameters provided above, we generate a \texttt{DUSTY} dust emission model of the $\sim10^{-6}$ M$_\odot$ dust component. The amplitude of the dust emission SED output from \texttt{DUSTY} is appropriately scaled by normalizing the total integrated stellar emission to $0.5$ L$_\odot$ at a distance of 4 kpc to \WT. As shown in Fig.~\ref{fig:320sed}, the projected emission from the $\sim10^{-6}$ M$_\odot$ dust component is lower than the dereddened emission covered by the JWST MIRI/LRS observations. 
The observational constraints from JWST therefore indicate that in addition to the warmer ($T_\mathrm{d}=720$ K) and less massive (M$_\mathrm{d}\sim10^{-11}$ M$_\odot$) dust component revealed from the +830 d SED modeling, up to $\sim10^{-6}$ M$_\odot$ of cool dust formed in the \WT~ejecta may also be present in the circumstellar environment of the remnant star.

In Sec.~\ref{sec:ap_320}, we conduct a similar analysis using the +320 d circumstellar dust parameters derived by \citet{De2023} and conclude that the parameters from our +320 d best-fit \texttt{DUSTY} model are more consistent with the observational constraints.

\subsubsection{Interpreting the Warm Dust Component as Ejecta Fallback}
\label{sec:fallback}
\begin{figure*}[t]
    \includegraphics[width=0.99\linewidth]{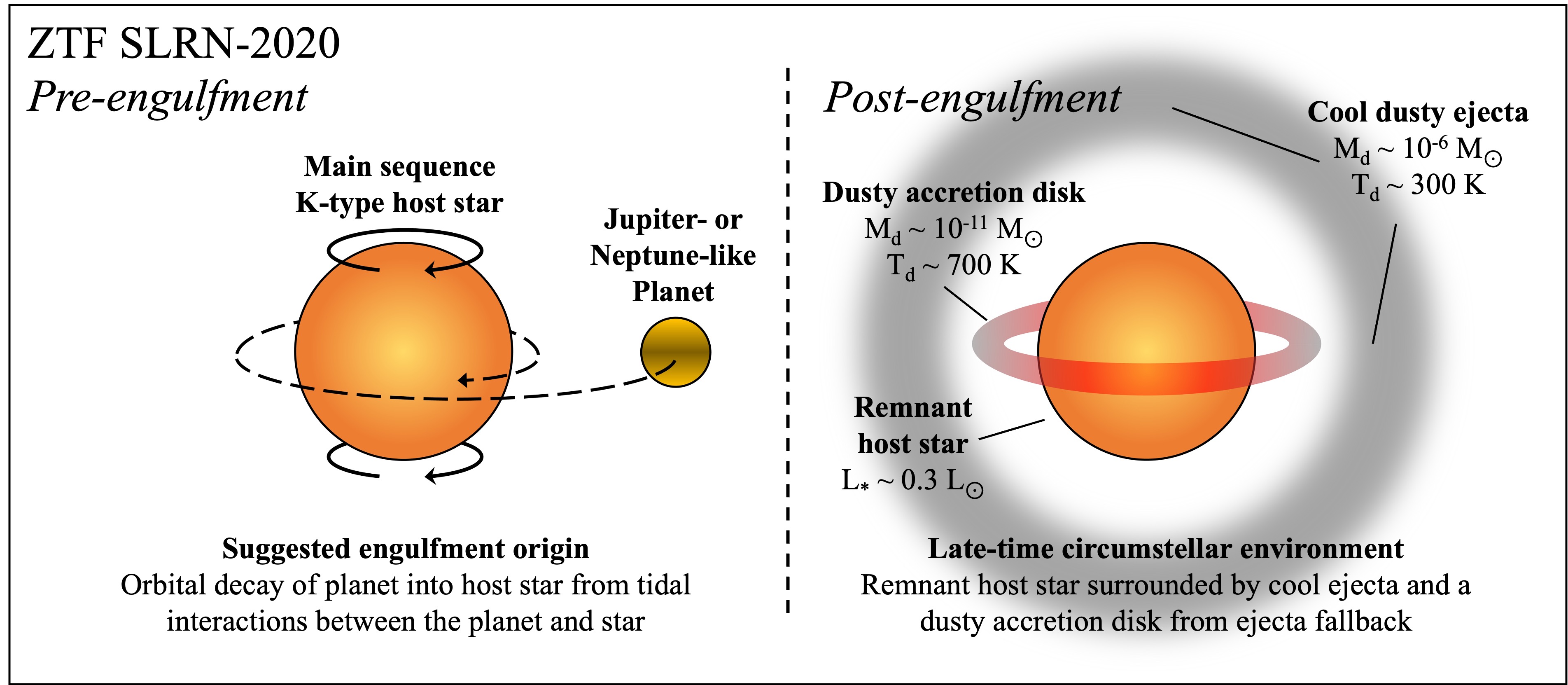}
    \caption{Schematic illustration of the pre- and post-engulfment interpretation of \WT. (\textit{Left}) The $\sim0.3$ L$_\odot$ luminosity of the \WT~remnant host star from \texttt{DUSTY} SED modeling (Sec.~\ref{sec:830}, Tab.~\ref{tab:Results}) is consistent with a K-type main sequence star. \WT~was therefore unlikely triggered by radial expansion from stellar evolution (Sec.~\ref{sec:nature}). Engulfment may instead be triggered by orbital decay of planet into the host from tidal interactions (Sec.~\ref{sec:decay}). (\textit{Right}) 
    At late times (+830 d), the remnant host star \RML{may be} surrounded by dusty accretion disk composed of fallback from the ejecta (Sec.~\ref{sec:fallback}). Ejecta dust detected by previous observations at +320 d (Sec.~\ref{sec:320}) is not detected by JWST at +830 d because the ejecta dust may have cooled beyond detection limits and wavelength coverage of observations (Sec.~\ref{sec:constraints}). } 
    \label{fig:summary}
\end{figure*}

We suggest that the warm, $\sim10^{-11}$ M$_\odot$ dust component revealed by JWST may have originated from
fallback from part of the ejecta (e.g.~\citealt{MacLeod2018b}). Dust condensed from the ejecta is likely traced by the cooler $\sim10^{-6}$ M$_\odot$ dust component observed at +320 d\rev{, especially given the evolution of this dust component throughout the \WT~outburst reported by \citet{De2023}.} 
The fallback interpretation is plausible given that the inferred ejecta velocity ($\sim100$ km s$^{-1}$) is less than the estimated escape velocity of $v_\mathrm{esc}\sim300$ km s$^{-1}$ \citep{De2023}. Given the detections of $^{12}$CO and Br$\alpha$ emission (Fig.~\ref{fig:SED}), the warm \RML{and dusty} circumstellar material may also \RML{be configured in an accretion disk around the remnant star as shown in Fig.~\ref{fig:summary} (\textit{Right})}.
\rev{Another possible interpretation of the warm $\sim10^{11}$ M$_\odot$ dust component may by a super-Eddington wind driven by energy deposition near the surface of the host star from the planetary engulfment (e.g.\citealt{Quataert2016}). However, \citet{De2023} measure a peak luminosity from \WT~($\sim10^{35}$ erg s$^{-1}$) that is well below the Eddington luminosity for a $\sim1$ M$_\odot$ star.}

We can investigate the ejecta fallback interpretation for the warm circumstellar dust component assuming the observed Br$\alpha$ emission (Fig.~\ref{fig:Bra}) arises from accretion of fallback ejecta on the remnant star. We assess whether or not the warm dust component, which has a total circumstellar mass of $M_\mathrm{CSM}\sim10^{-9}$ M$_\odot$ (Eq.~\ref{eq:warm_csm}),  provides a sufficient mass reservoir to power the observed Br$\alpha$ line luminosity. 

Based on the Br$\alpha$ line flux from the \rev{Gaussian} profile fit (Tab.~\ref{tab:Bralpha}) and adopting a distance of 4 kpc to \WT, the Br$\alpha$ luminosity is 

\begin{equation}
    \rev{\mathrm{L}_{\mathrm{Br}\alpha}\sim10^{-5}\, \mathrm{L}_\odot}
\end{equation}

\noindent
The luminosity of hydrogen recombination lines have notably been used as proxies for accretion luminosity for systems such as T Tauri stars (e.g.~\citealt{Salyk2013}), a class of young and accreting low-mass (M$_*<$ 2 M$_\odot$) stars. The hydrogen emission lines from T Tauri stars are thought to originate from the accretion columns and accretion shock \citep{Calvet1998}. Assuming a similar mechanism is driving the Br$\alpha$ emission for the \WT~remnant star, we estimate L$_\mathrm{acc}$ from L$_{\mathrm{Br}\alpha}$ by adopting the empirical L$_{\mathrm{Br}\alpha}$-L$_\mathrm{acc}$ relation derived for T Tauri stars \citep{Komarova2020}:

\begin{equation}
    \mathrm{Log}\left(\frac{\mathrm{L}_\mathrm{acc}}{\mathrm{L}_\odot}\right)= (1.81\pm0.11)\,\mathrm{Log}\left(\frac{\mathrm{L}_{\mathrm{Br}\alpha}}{\mathrm{L}_\odot}\right)+(6.45\pm0.38).
\label{eq:Lacc}
\end{equation}

Based on Eq.~\ref{eq:Lacc}, we infer an accretion luminosity for the remnant star of \WT,

\begin{equation}
    \rev{\mathrm{L}_\mathrm{acc}\sim4\times10^{-3}\,\mathrm{L}_\odot.}
\end{equation}

\noindent
From the accretion luminosity, we can approximate the remnant star's mass accretion rate, $\dot{M}_\mathrm{acc}$, given the following relation

\begin{equation}
    L_\mathrm{acc} \approx G\, M_*\,\dot{M}_\mathrm{acc} / R_*,
\end{equation}

\noindent
where M$_*$ is the mass of the remnant star, R$_*$ is the radius of the remnant star, and $G$ is the gravitational constant. Assuming $\frac{\mathrm{M}_*}{\mathrm{R}_*}\sim\frac{\mathrm{M}_\odot}{\mathrm{R}_\odot}$, we estimate a mass accretion rate of

\begin{equation}
    \rev{\dot{M}_\mathrm{acc}\sim10^{-10}\,\mathrm{M}_\odot\, \mathrm{yr}^{-1}.}
\end{equation}

\noindent
The $\sim10^{-9}$ M$_\odot$ of warm circumstellar material \rev{(Eq.~\ref{eq:warm_csm})} indeed provides a sufficient mass reservoir to power the estimated mass accretion rate and therefore supports the interpretation of the warm dust component as fallback from the ejecta.





\subsection{Addressing the Nature of \WT}
\label{sec:nature}


Our JWST results and analysis indicate that the luminosity of the \WT~remnant host star is consistent with a K-type main-sequence star even if it were obscured by 10$^{-6}$ M$_\odot$ of ejecta dust (Sec.~\ref{sec:830} \&~\ref{sec:obscured}). The properties of the remnant star should reflect that of the progenitor since the remnant star has likely relaxed to its normal state based on the stability of its near-IR emission (Fig.~\ref{fig:SED}). \WT~was therefore unlikely triggered by radial expansion from stellar evolution since the main sequence lifetime for K-type stars ($\gtrsim15$ Gyr) is longer than the age of the Universe. 
In this section, we discuss the implications of our results and the persisting questions on the nature of \WT~planetary engulfment event. 

We raise an important caveat on the measured stellar luminosity arising from the uncertainty on the estimated distance to \WT. Based on an a comparative analysis of extinction maps, \citet{De2023} claim $d=4$ kpc is the best distance estimate for the source but indicate a conservative distance range of 2 -- 7 kpc. Given our L$_*\approx 0.3$ L$_\odot$ calculation for the host star luminosity (Tab.~\ref{tab:Results}), the upper-bound distance would imply a stellar luminosity and mass of $\approx0.9$ L$_\odot$ and $\approx1$ M$_\odot$, respectively, and such a star may have been evolving off of the main sequence. 
However, the three Galactic dust extinction maps \citep{Drimmel2003,Marshall2006,Green2019} used by \citet{De2023} to estimate the distance show consistent overlaps between the range of $\approx3$ -- 4 kpc, which motivated adopting 4 kpc as the best distance estimate towards \WT.

\subsubsection{Engulfment triggered by orbital decay via tidal interactions?}
\label{sec:decay}
If \WT~was not caused by stellar evolution, a plausible hypothesis for the mechanism that triggered the planetary engulfment event is tidal interactions between the host star and the planet companion \RML{(See Fig.~\ref{fig:summary}, \textit{Left})}. Such interactions can lead to an orbital decay of the planet into the host star as energy and angular momentum are transferred from the orbit of the planet to the host star (e.g. \citealt{Zahn1977,Jackson2009}). Evidence of orbital decay from tidal interactions has notably been provided from transit-timing observations of the hot Jupiters WASP-12b \citep{Maciejewski2016,Patra2020,Yee2020} and Kepler-1658b \citep{Vissapragada2022}. \citet{Hamer2019} also argue that hot Jupiters are destroyed by tides during the main-sequence lifetimes of their host stars based on a Galactic velocity dispersion study of stars with and without hot Jupiters. The orbital decay timescales based on recent simulations for host stars $\lesssim1$ M$_\odot$ and planets with an orbital period $\lesssim1$ day are $<10^8$ yr \citep{Weinberg2024}, which is notably much shorter than the main-sequence lifetime of a K-type star.


Other mechanisms that could lead to planetary engulfment include planet-planet scattering \citep{Rasio1996,Chatterjee2008,Carrera2019} and gravitational perturbations from another companion \citep{Holman1997, Naoz2012, Stephan2018}. These mechanisms could excite high orbital eccentricities such that interactions between the planet and the host star might occur at periapse, 
though \citet{De2023} argue that \WT~must have had a nearly-circular orbit by the time the planet was engulfed due to the steadily rising pre-outburst luminosity.
We therefore favor the orbital decay scenario. However, observations of additional planetary engulfment events will be crucial for assessing the dominant mechanism(s).

\subsubsection{Engulfment or Tidal Disruption?}
\label{Sec:etd}
Understanding the nature of the host star can provide insight into the physics behind \WT. In particular, the ratio of the mean densities of the planet and host star is thought to be a proxy for the outcome of the star-planet interaction \citep{Metzger2012}.
A K-type main sequence star has a mean stellar density of $\bar{\rho}_*\sim3$ g cm$^{-3}$, which is greater than that of hot Jupiters that tend to be ``inflated'' \citep{Baraffe2010} and exhibit relatively low mean densities ($\bar{\rho}_p\lesssim$ 1 g cm$^{-3}$).
Instead of being fully engulfed, which should occur when the mean density of the planet is greater than that of the host star, a planet with a mean density lower than that of their host star may undergo tidal disruption near or above the stellar surface. Tidal disruption of the planet would lead to the formation of an accretion disk that can drive outflows from super-Eddington accretion onto the host star \citep{Metzger2012,Matsumoto2022}. 

Alternatively, given the Neptune-like ($\sim0.1$ M$_\mathrm{Jup}$) lower-end mass estimate of the engulfed companion by \citet{De2023}\footnote{The $0.1$ M$_\mathrm{Jup}$ mass estimate was derived from linearly scaling the companion mass in the stellar merger V1309 Sco \citep{Tylenda2011} based on the ejecta mass and radiated energy of V1309 Sco and \WT.},
we speculate that perhaps the engulfed planet from \WT~was in the ``Hot Neptune Desert" \citep{Szabo2011}. Among the handful of known planets in the Hot Neptune Desert, several exhibit mean densities of $\bar{\rho}_p\gtrsim3$ g cm$^{-3}$ (e.g.~\citealt{Armstrong2020,Naponiello2023,Nabbie2024}). A ``hot Neptune" could therefore be engulfed by a K-type main-sequence host star. However, since such planets appear to be relatively rare compared to hot Jupiters, the star-planet interaction leading to \WT~more likely involved a hot Jupiter than a hot Neptune. 


The physics that ultimately drive the mass ejection in tidal disruption scenario is distinct from the engulfment scenario where the planet deposits its orbital energy into its surroundings as it plunges into the host star \citep{Macleod2018a}. 
Signatures of chemical enrichment of the host star from the planet may differ for each scenario since ejecta from a merger/engulfment should be dominated by material from the host star envelope \citep{Metzger2012,Matsumoto2022}, whereas the accretion disk from tidal disruption will be composed of the planetary material \citep{Metzger2012}.
Both scenarios, however, can result in a transient outburst with potentially similar observational properties (e.g.~\citealt{Bear2011,Soker2023}). 
Spectroscopic follow-up of \WT~searching for signatures of chemical enrichment may help distinguish its origin as well as provide a valuable dataset for investigating anomalous stellar chemical compositions, which indicate at least one in a dozen stars presents evidence of planet engulfment \citep{Liu2024}. 

Given the similarities of \WT~to other engulfment/merger-powered red novae \citep{De2023}, such as the presence of multiple dust components (e.g.~\citealt{Tylenda2016,Woodward2021}), we favor the engulfment interpretation for \WT. We also note that although hot Jupiters tend to possess relatively low mean densities, hot Jupiters with mean densities $\bar{\rho}_p\gtrsim3$ g cm$^{-3}$ are known to exist (e.g.~\citealt{Deleuil2012,Rouan2012,Espinoza2017}). 
We acknowledge that the $\bar{\rho}_p/\bar{\rho}_*<1$ mean density ratio between a K-type star and a typical hot Jupiter-like planet as well as the deviation from the $L_*\propto t^{-4/5}$ power-law decay expected for a merger/engulfment event (Sec.~\ref{sec:320}) present open questions on the engulfment interpretation.
Further theoretical/simulation work on star-planet interactions (e.g.~\citealt{Lau2022,Yarza2023}) in addition to multi-wavelength observations of more subluminous red novae like \WT~will be important for distinguishing the physics that drive these transient events.

\section{Summary and Conclusions}


In this paper, we presented late-time (+830 d) spectroscopic IR observations of the planetary engulfment event \WT~with JWST/NIRSpec and MIRI. The $\sim3-12$ $\mu$m spectra from JWST were complemented with near-contemporaneous, near-IR photometry from ground-based imaging with Gemini-N/NIRI (Fig.~\ref{fig:SED}). In our analysis of the $\sim1-12$ $\mu$m SED of \WT, we also incorporated previously published near-IR photometry from P200/WIRC taken a few weeks before the Gemini-N observations \citep{De2023}. A schematic illustration of the pre- and post-engulfment interpretation of \WT~that summarizes the main results and implications is shown in Figure~\ref{fig:summary}).

The $\sim1-12$ $\mu$m observations of \WT~likely trace emission from the remnant host star in the near-IR and thermal dust emission at longer wavelengths in the mid-IR. We identified the following notable properties from the dereddened $\sim1-12$ $\mu$m SED: a peak in the near-IR around the $H$-band, the detection of Br$\alpha$ emission, \rev{the potential detection of PH$_3$ emission at $\sim4.3$ $\mu$m}, and the presence of $^{12}$CO fundamental band emission at $\sim4.7$ $\mu$m. Interestingly, $^{12}$CO emission was also detected around the red nova V4332 Sgr \citep{Banerjee2004}, which has been compared to \WT~due to its low luminosity \citep{De2023}. 
The detections of $^{12}$CO fundamental band and Br$\alpha$ emission suggest the presence of hot circumstellar gas that may be accreting onto the host star.

We modeled the $^{12}$CO emission from \WT~using \texttt{spectools\_ir} which revealed hot (\rev{T$_\mathrm{CO}\sim1340$ K}) and close-in (\rev{R$_\mathrm{CO}\sim15$ R$_\odot$}) circumstellar gas with a total CO mass of \rev{$\mathrm{Log}\left(\frac{M_\mathrm{CO}}{\mathrm{M}_\odot}\right)\approx-13.1^{+0.3}_{-0.3}$} (Fig.~\ref{fig:CO}, Tab.~\ref{tab:Results}). Based on the CO gas mass, we estimated a total gas mass of \rev{$\mathrm{Log}\left(\frac{M_\mathrm{gas}}{\mathrm{M}_\odot}\right)\approx-10.2^{+0.3}_{-0.3}$}.
Although the $^{12}$CO fundamental band in emission is typically detected from YSOs, Herbig Ae/Be stars, and T Tauri disks, the observed properties of \WT~presented in this work and by \citet{De2023} conflict with these interpretations.
The JWST spectra are therefore consistent with the claim that \WT~arose from a planetary engulfment event.

\rev{We analyzed the broad emission feature at $\sim4.3$ $\mu$m with synthetic spectral models from iSLAT and attributed the feature to PH$_3$. The analysis suggests the presence of $\sim2\times10^{-11}$ M$_\odot$ of 300 K PH$_3$ assuming it is co-spatial with the circumstellar material at a distance of 22 au from the remnant host star. The abundance of PH$_3$ relative to H$_2$ is approximately $\sim10^{-8}$, which is notably consistent with that of the circumstellar material around the AGB star IRC +10216 \citep{Agundez2014}.}

We investigated the properties of the \WT~remnant host star and circumstellar dust by fitting \texttt{DUSTY} radiative transfer models. Based on the best-fit model to the late-time (+830 d) $1-12$ $\mu$m SED, we derived a luminosity of L$_*\approx 0.3$ L$_\odot$ for the surviving host star, which is consistent with a $\sim0.7$ M$_\odot$ K-type star. 
From the late-time, +830 d SED, we derived a circumstellar dust mass of \rev{$\mathrm{Log}\left(\frac{M_\mathrm{d}}{\mathrm{M}_\odot}\right) \approx -10.61^{+0.08}_{-0.16}$} and dust temperature of $T_\mathrm{d}=720$ K. Assuming a gas-to-dust mass ratio of 100, the we estimated a total circumstellar mass of \rev{$\mathrm{Log}\left(\frac{M_\mathrm{CSM}}{\mathrm{M}_\odot}\right) \approx -8.61^{+0.08}_{-0.16}$}. The circumstellar dust component revealed from the +830 d SED may therefore be distinct from the cooler and more massive component probed by the +320 d SED from \citet{De2023}. In order to investigate this, we re-analyzed the +320 d SED using a similar grid-search approach as the +830 d SED analysis with \texttt{DUSTY} (Fig.~\ref{fig:320sed}). Our revised analysis of the +320 d SED provided a total dust mass of Log$\left(\frac{M_\mathrm{d}}{\mathrm{M}_\odot}\right)=-5.89^{+0.29}_{-3.21}$ and dust temperature of $T_\mathrm{d}\sim280$ K (Tab.~\ref{tab:Results}). We therefore suggested that the remnant star of \WT~ has at least two distinct circumstellar dust components.

Given the likely presence of a cool, massive circumstellar dust shell around the remnant star of \WT, which was not revealed by the best-fit +830 d SED model, we addressed:

\begin{itemize}
\itemsep0em 
    \item How might this impact our +830 d SED model hypothesis that \WT~hosts a $\sim0.7$ M$_\odot$ K-type star.
    \item Whether such a massive but cool circumstellar dust component could have been missed by the JWST observations. 
\end{itemize}

\noindent


When factoring in the obscuration from a projected $\sim10^{-6}$ M$_\odot$ dust component at the +830 d epoch, our analysis indicated that the dust-corrected luminosity of the remnant star would still be consistent with a K-type star with a mass $\lesssim0.85$ M$_\odot$. Given that such a star should not have evolved off of the main sequence, we suggest that the planetary engulfment event associated with \WT~was not likely triggered by radial expansion from stellar evolution. 

In order to address whether the JWST could have missed a cool circumstellar dust component, we modeled the predicted emission from the $\sim10^{-6}$ M$_\odot$ dust component at the +830 d epoch. As shown in Fig.~\ref{fig:320sed}, we determined that $\sim10^{-6}$ M$_\odot$ of circumstellar dust formed in the \WT~ejecta could indeed have been undetected by the +830 d JWST observations. Given that the cooler and more massive dust component likely condensed in the ejecta of \WT~\citep{De2023}, we hypothesize that the warmer dust component revealed by the JWST observations traces fallback from the ejecta and is being accreted on the remnant host star.
In support of this hypothesis, we calculated that the $\sim10^{-9}$ M$_\odot$ of total material in the warm circumstellar component can provide a sufficient mass reservoir to power the \rev{$\sim10^{-10}$ M$_\odot$ yr$^{-1}$} mass accretion rate derived from the Br$\alpha$ emission line luminosity.

If \WT~was not triggered by stellar evolution, we favor the hypothesis that the planetary engulfment was due to orbital decay caused by tidal interactions between the planet and the host star. Other possible mechanisms include planet-planet scattering and gravitational perturbations from another companion. However, those two mechanisms excite high orbital eccentricities, which is not consistent with the nearly-circular orbit inferred for the system by \citet{De2023}.

Lastly, we considered the implications on the star-planet interaction leading to \WT~if the host is a main-sequence K-type star. Since the mean density of the star would be greater than that of a hot Jupiter-like planet, the planet may be tidally disrupted near or above the surface of the star rather than being fully engulfed. Noting that a Neptune-like $\sim0.1$ M$_\mathrm{Jup}$ planet was also consistent for the engulfed companion \citep{De2023}, we speculated that the planet may have been a Hot Neptune, which exhibit higher mean densities than Hot Jupiters. Ultimately, we continue to favor the engulfment interpretation for \WT. The discovery and follow-up of new planetary engulfment events will be important for investigating the physics of these IR-luminous transient events. This study of \WT~highlights the scientific potential of coordinated IR-bright transient discovery and follow-up observations with JWST and upcoming survey facilities like the Vera C.~Rubin Observatory.





\begin{acknowledgments}
We thank our JWST program coordinator Shelly Meyett for facilitating our Target of Opportunity (ToO) Activation Request, which was one of the first ToO triggers with JWST.  
We also thank Sarah Kendrew for the valuable feedback on the reduction and analysis of the MIRI LRS observations. We also thank Greg Sloan for his review and feedback on our offset target acquisition strategy for the MIRI LRS observations.
RML would like to acknowledge Joan Najita, J.J.~Zanazzi, and Viraj Karambelkar for enlightening discussions on circumstellar disks, planetary engulfment events, and stellar mergers. RML also thanks Sarah Logsdon \rev{and Everett Schlawin} for the helpful discussions on exoplanets.


This work is based on observations made with the NASA/ESA/CSA James Webb Space Telescope. The data were obtained from the Mikulski Archive for Space Telescopes at the Space Telescope Science Institute, which is operated by the Association of Universities for Research in Astronomy, Inc., under NASA contract NAS 5-03127 for JWST. These observations are associated with program \#1240.

\rev{Some of the data presented in this article were obtained from the Mikulski Archive for Space Telescopes (MAST) at the Space Telescope Science Institute. The specific observations analyzed can be accessed via \dataset[10.17909/198x-0r66]{https://doi.org/10.17909/198x-0r66}}

Based on observations obtained at the international Gemini Observatory, a program of NSF NOIRLab, which is managed by the Association of Universities for Research in Astronomy (AURA) under a cooperative agreement with the U.S. National Science Foundation on behalf of the Gemini Observatory partnership: the U.S. National Science Foundation (United States), National Research Council (Canada), Agencia Nacional de Investigaci\'{o}n y Desarrollo (Chile), Ministerio de Ciencia, Tecnolog\'{i}a e Innovaci\'{o}n (Argentina), Minist\'{e}rio da Ci\^{e}ncia, Tecnologia, Inova\c{c}\~{o}es e Comunica\c{c}\~{o}es (Brazil), and Korea Astronomy and Space Science Institute (Republic of Korea). 
The work of RML is supported by NOIRLab.

K. D. was supported by NASA through the NASA Hubble Fellowship grant $\#$HST-HF2-51477.001 awarded by the Space Telescope Science Institute, which is operated by the Association of Universities for Research in Astronomy, Inc., for NASA, under contract NAS5-26555.
M.M. is grateful for support from a Clay Postdoctoral Fellowship at the Smithsonian Astrophysical Observatory.

\end{acknowledgments}

\vspace{5mm}
\facilities{JWST/MIRI, JWST/NIRSpec, Gemini-N/NIRI}





\appendix

\section{Circumstellar Dust Mass Calculation from \texttt{DUSTY} Parameters}
\label{sec:appendix}

We demonstrate how the circumstellar dust can be estimated from the \texttt{DUSTY} SED models parameters using the \texttt{SPHERE} geometry, which assumes spherical symmetry in density. The optical depth due to the circumstellar dust, $\tau_V$, which is one of the free parameters in the \texttt{DUSTY} models, can be defined as follows:


\begin{equation}
    \tau_V = \int_{R_\mathrm{in}}^{R_\mathrm{out}}\rho(r)\kappa_V^{d} \,\mathrm{d}r,
    \label{eq:tau}
\end{equation}

\noindent
where $\kappa_V^d$ is the dust opacity, defined here as the dust absorption cross section weighted by dust mass\footnote{Note that dust opacity may also be defined as the dust absorption cross section weighted by dust \textit{and gas} mass (e.g.~\citealt{De2023})}, and $\rho(r)$ is the dust density as a function of radius, $r$, from the center of the circumstellar shell. We define the dust density for the spherical circumstellar shell as

\begin{equation}
    \rho(r)=\rho_0\left(\frac{r}{R_\mathrm{in}}\right)^{-\alpha},
    \label{eq:density}
\end{equation}

\noindent
where $\rho_0$ is the dust density at the inner radius of the dust shell, $R_\mathrm{in}$, and $\alpha$ is the power-law index of the radial dust density profile. The dust mass, $M_d$, is expressed simply by the integral of the dust density over the volume of the circumstellar dust shell with inner radius, $R_\mathrm{in}$, and outer radius, $R_\mathrm{out}$:

\begin{equation}
    M_{d} = 4 \pi \int_{R_\mathrm{in}}^{R_\mathrm{out}}r^2 \rho(r) \,\mathrm{d}r.
    \label{eq:dust}
\end{equation}

\noindent
By combining Equations~\ref{eq:tau},~\ref{eq:density}, and~\ref{eq:dust}, and substituting $R_\mathrm{out}$ for the shell thickness parameter $Y= R_\mathrm{out}/R_\mathrm{in}$, which is an input parameter from \texttt{DUSTY}, the dust mass can be expressed as:

\begin{equation}
\begin{aligned}
    M_{d} = \frac{4 \pi \tau_V}{\kappa_V^{d}} \frac{\int_{R_\mathrm{in}}^{R_\mathrm{out}}r^{2-\alpha} \,\mathrm{d}r}{\int_{R_\mathrm{in}}^{R_\mathrm{out}}r^{-\alpha} \,\mathrm{d}r} \\
    =  \frac{4 \pi \tau_V R_\mathrm{in}^2}{\kappa_V^{d}}\frac{(\alpha -1) \left(Y^{\alpha }-Y^3\right)}{(\alpha -3) \left(Y^{\alpha }-Y\right)}.
    \end{aligned}
    \label{eq:mass2}
\end{equation}

\noindent
Assuming a radial density power-law index of $\alpha=2$, the dust mass of the circumstellar shell is 

\begin{equation}
    M_{d} = \frac{4 \pi \tau_V R_\mathrm{in}^2 Y }{\kappa_V^{d}}, 
    \label{eq:mass3}
\end{equation}

\noindent
Note that $R_\mathrm{in}$ is a parameter output from the \texttt{DUSTY} modeling.

As an example, in Sec.~\ref{sec:830}, the \WT~dust mass calculated at the +830 d epoch from with the best-fit \texttt{DUSTY} parameters (Tab.~\ref{tab:Results}) is calculated as follows:

\begin{equation}
    \frac{M_{d}}{M_\odot} \approx 2.4 \times10^{-11} \left(\frac{R_\mathrm{in}}{50\, R_\odot}\right)^2 \left(\frac{\tau_V}{0.7}\right) \left(\frac{Y}{5}\right) \left(\frac{\kappa_V^{d}}{1.123\times10^4\,\mathrm{cm}^2\,\mathrm{g}^{-1}}\right)^{-1}. 
\end{equation}


\section{Equilibrium Dust Temperature from a Central Stellar Heating Source}
\label{sec:teq}

The balance between the input radiation from a central heating source with luminosity L$_*$ and the thermal emission from circumstellar dust at a temperature T$_\mathrm{eq}$ can be expressed as follows

\begin{equation}
    \frac{L_*}{4\pi r^2}\left<Q_\mathrm{abs}\right>_*= 4 \left<Q_\mathrm{abs}\right>_{T_\mathrm{eq}} \sigma_\mathrm{SB} T_\mathrm{eq}^4, 
\end{equation}


\noindent
where $r$ is the separation distance between dust and the heating source, $\sigma_\mathrm{SB}$ is the 
Stefan–Boltzmann constant, $\left<Q_\mathrm{abs}\right>_*$ is the spectrum-averaged absorption cross section from the heating source, and $\left<Q_\mathrm{abs}\right>_{T_\mathrm{eq}}$ is the Planck-averaged emission efficiency from the emitting dust (See \citealt{Draine2011}).

Assuming the radiation from the heating source can be characterized by a blackbody function at temperature $T_*$, the spectrum-averaged absorption cross section is defined as 

\begin{equation}
\left<Q_\mathrm{abs}\right>_*=\frac{\int B_\nu(T_*) Q_\mathrm{abs}(\nu) d \nu}{\int B_\nu(T_*) d \nu},
\label{eq:Qstar}
\end{equation}
\noindent
where $Q_\mathrm{abs}(\nu)$ is the dust absorption efficiency. The Planck-averaged emission efficiency is defined as

\begin{equation}
\left<Q_\mathrm{abs}\right>_{T_\mathrm{eq}}=\frac{\int B_\nu(T_\mathrm{eq}) Q_\mathrm{abs}(\nu) d \nu}{\int B_\nu(T_\mathrm{eq}) d \nu}.
\end{equation}

\noindent
The dust absorption efficiency $Q_\mathrm{abs}(\nu)$ is commonly approximated as a power-law in frequency/wavelength with index $\beta$ ( $Q_\mathrm{abs}(\nu)\propto\nu^\beta$), which is typically assumed to be $0<\beta<2$. By adopting the power-law approximation for $Q_\mathrm{abs}(\nu)$, it follows that $\left<Q_\mathrm{abs}\right>_{T_\mathrm{eq}}\propto T_\mathrm{eq}^\beta$. 

Assuming that the dust grain size distribution and composition is fixed, the relation between the dust equilibrium temperature and the heating source parameters can be expressed as


\begin{equation}
T_\mathrm{eq}\propto \left<Q_\mathrm{abs}\right>_*^{1/(4+\beta)}  L_*^{1/(4+\beta)} r^{-2/(4+\beta)}.
\label{eq:teq}
\end{equation}

\section{Constraints on Previous +320 d SED Model Parameters}
\label{sec:ap_320}
We conducted a similar analysis using the +320 d circumstellar dust parameters derived by \citet{De2023}: $\tau_V=13$, $T_d=415$ K, T$_*=4300$ K, and R$_\mathrm{in}=1143$ R$_\odot$. The projected circumstellar dust component at +830 d based on the \citet{De2023} parameters would exhibit a 10 $\mu$m flux of brightness of $\sim10^{-3}$ Jy, an order of magnitude greater than the 10 $\mu$m flux measured by the JWST observations at +830 d.  

The difference in the $\sim10$ $\mu$m brightness between the projected dust component from the \citet{De2023} +320 d parameters and ours is due to the differing inner radii (1143 R$_\odot$ vs 4730 R$_\odot$), which imply slower moving material for the \citet{De2023} dust component ($\sim30$ km s$^{-1}$ vs $\sim120$ km s$^{-1}$). The inner radius and dust temperature from the \citet{De2023} +320 d model are notably consistent with the uncertainties of our +320 d fit (Tab.~\ref{tab:Results}). The projected +830 d dust temperature from the \citet{De2023} +320 d component is therefore hotter and thus brighter at $\sim10$ $\mu$m than the projected component from our +320 d model ($T_\mathrm{d}^\mathrm{+830d}\approx 205$ K vs 100 K). Based on our projected dust emission analysis and the constraints from the JWST observations, we favor the parameters from our +320 d best-fit \texttt{DUSTY} model for the ejecta dust.





\bibliography{mybib}{}
\bibliographystyle{aasjournal}



\end{document}